\documentclass[aps,pre,twocolumn,floatfix]{revtex4}%
\usepackage{amsfonts}
\usepackage{amsmath}
\usepackage{amssymb}
\usepackage{graphicx}
\usepackage{bm}
\usepackage{subfigure}
\usepackage{color}
\usepackage[colorlinks,linkcolor=blue, urlcolor=blue,
anchorcolor=blue,citecolor=blue,breaklinks=true]{hyperref}%
\providecommand{\U}[1]{\protect\rule{.1in}{.1in}}
\providecommand{\U}[1]{\protect\rule{.1in}{.1in}}
\begin{document}
\title{Quantum Hall effect in topological Dirac semimetals modulated by the Lifshitz
transition of the Fermi arc surface states}
\author{Tao-Rui Qin$^{1}$}
\author{Zhuo-Hua Chen$^{1}$}
\author{Tian-Xing Liu$^{1}$}
\author{Fu-Yang Chen$^{1}$}
\author{Hou-Jian Duan$^{1,2}$}
\author{Ming-Xun Deng$^{1,2}$}
\email{dengmingxun@scnu.edu.cn}
\author{Rui-Qiang Wang$^{1,2}$}
\email{wangruiqiang@m.scnu.edu.cn}
\affiliation{$^{1}$Guangdong Provincial Key Laboratory of Quantum Engineering and Quantum
Materials, School of Physics, South China Normal University, Guangzhou 510006, China}
\affiliation{$^{2}$Guangdong-Hong Kong Joint Laboratory of Quantum Matter, Frontier
Research Institute for Physics, South China Normal University, Guangzhou
510006, China}

\begin{abstract}
We investigate the magnetotransport of topological Dirac semimetals (DSMs) by
taking into account the Lifshitz transition of the Fermi arc surface states.
We demonstrate that a bulk momentum-dependent gap term, which is usually
neglected in study of the bulk energy-band topology, can cause the Lifshitz
transition by developing an additional Dirac cone for the surface to prevent
the Fermi arcs from connecting the bulk Dirac points. As a result, the Weyl
orbits can be turned off by the surface Dirac cone without destroying the bulk
Dirac points. In response to the surface Lifshitz transition, the Weyl-orbit mechanism for the 3D quantum
Hall effect (QHE) in topological DSMs will break down. The resulting quantized Hall plateaus can be thickness-dependent, similar to the Weyl-orbit
mechanism, but their widths and quantized values become irregular. Accordingly, we propose that apart from the bulk Weyl nodes and Fermi arcs, the
surface Lifshitz transition is also crucial for realizing stable Weyl orbits and 3D QHE in real materials.

\end{abstract}
\maketitle

\section{introduction}

Topological semimetals are novel quantum states of matter, in which the
conduction and valence bands cross near the Fermi level at certain discrete
momentum points or
lines\cite{RevModPhys.90.015001,Neupane:2014aa,Cd3As2_PhysRevB.88.125427,A3B_PhysRevB.85.195320,Liu864,Yang:2014aa,PhysRevB.83.205101,Zhang:2016aa,Zhang:2017aa,PhysRevLett.119.136806}%
. The gap-closing points or lines are protected either by crystalline symmetry
or topological invariants\cite{PhysRevB.91.121101,Kargarian8648}. A
topological Dirac semimetal (DSM) hosts paired gap-closing points, referred to
as the Dirac points, which are stabilized by the time-reversal,
spatial-inversion and crystalline symmetries. By breaking the time-reversal or
spatial-inversion symmetry, a single Dirac point can split into a pair of Weyl
nodes of opposite chiralities\cite{NIELSEN1983389,Volovik2003}, leading to the
topological transition from a Dirac to a Weyl
semimetal\cite{PhysRevX.9.011039,PhysRevB.85.165110,PhysRevB.88.245107,PhysRevB.98.085149,PhysRevB.96.155141,PhysRevB.99.075131}%
. Accompanied with the bulk topological transition, topological states which
are protected by the quantized Chern flux will emerge in the surface to
connect the split Weyl nodes, known as the Fermi-arc surface
states\cite{PhysRevB.83.205101}.

In topological DSMs, such as \textrm{A}$_{3}$\textrm{Bi} (\textrm{A} =
\textrm{Na},\textrm{K},\textrm{Rb})\cite{A3B_PhysRevB.85.195320,Liu864} and
\textrm{Cd}$_{2}$\textrm{As}$_{3}$%
\cite{Cd3As2_PhysRevB.88.125427,Neupane:2014aa}, the Weyl nodes at the same
Dirac point, belonging to different irreducible representations, cannot be
coupled and have to seek for a partner from the other Dirac point. As a
consequence, the two Dirac points including two pairs of Weyl nodes are
connected by two spin-polarized Fermi
arcs\cite{A3B_PhysRevB.85.195320,Cd3As2_PhysRevB.88.125427,Liu864,Yang:2014aa,PhysRevB.83.205101}%
. The Fermi-arc surface states are the most distinctive observable
spectroscopic feature of topological semimetals. However, their observation is
sometime limited by spectroscopic resolutions. There have been searching for
alternative smoking-gun features of topological semimetals, such as by means
of transport phenomena\cite{Ali2014,Shekhar2015,PhysRevX.4.031035}. Many
interesting transport properties have been revealed in topological semimetals,
for example, chiral anomaly induced negative
magnetoresistance\cite{Xiong413,Li:2015aa,PhysRevLett.122.036601,PhysRevB.99.165146}%
, Weyl-orbit related quantum oscillations\cite{Nat.Commun.5.5161,Moll2016},
Berry's phase $\pi$ related Aharonov-Bohm
effect\cite{PhysRevB.95.235436,Wang20166}, bulk-surface interference induced
Fano effect\cite{PhysRevLett.120.257701}, topological pumping
effect\cite{PhysRevB.106.075139}, etc.

Recently, 3D quantum Hall effect (QHE) because of the Fermi arcs was proposed
in Weyl semimetals\cite{PhysRevLett.119.136806} and has led to an explosion of
theoretical\cite{PhysRevLett.125.036602,PhysRevB.104.075425,PhysRevB.104.205305,PhysRevB.103.245434,PhysRevB.103.245409,lu20193d,li20213d,PhysRevLett.125.206601,Gooth_2023,PhysRevB.107.125203,du2021quantum}
and
experimental\cite{uchida2017quantum,PhysRevLett.120.016801,zhang2019quantum,tang2019three,PhysRevLett.122.036602}
activities in the field of condensed matter physics. In a Weyl semimetal slab,
the Weyl orbit which consists of Fermi arcs from opposite surfaces can support
the electron cyclotron orbit for the QHE, making the 3D QHE available. The 3D
QHE has been observed experimentally in topological
DSMs\cite{uchida2017quantum,PhysRevLett.120.016801,zhang2019quantum,tang2019three,PhysRevLett.122.036602}
and the one from the Weyl-orbit mechanism is demonstrated to be
thickness-dependent \cite{zhang2019quantum}. However, for the topological
DSMs, a single surface with two Fermi arcs can also support a complete Fermi
loop required by the QHE, which can compete with the Weyl-orbit mechanism. The
same-surface Fermi loop is not stable and can be deformed by bulk
perturbations\cite{Kargarian8648}. In real materials, the bulk perturbations are inevitable.

As we will show, when a bulk
momentum-dependent gap term is included, Lifshitz transition can happen for
the Fermi arc surface states, in which the double Fermi arcs on the DSM
surface can be continuously deformed into a closed Fermi loop and separate
from the bulk Dirac points. The Lifshitz transition involves a change of the
Fermi surface topology that is not connected with a change in the symmetry of
the lattice\cite{lifshitz1960anomalies,science.1239451,Zeljkovic2014,PhysRevLett.115.166602,PhysRevB.104.155135}. Therefore, the Lifshitz transition can take place without destroying the topology of the bulk energy band. A natural question in this regard is how the deformation of the
Fermi arcs influences the 3D QHE of topological DSMs, especially when the Fermi arcs, as key ingredients for the Weyl orbits, breaks free from the bulk Dirac points.

In this paper, we investigate the QHE in topological DSMs by taking into
account the surface Lifshitz transition, which can be modulated by a bulk momentum-dependent gap term. It
is demonstrated that while the bulk Dirac points are robust against the
momentum-dependent gap term, the surface can develop an additional 2D Dirac
cone, which deforms the surface Fermi arcs from a curve to some discrete
points and further to a Fermi loop coexisting with the bulk Dirac points.
During this process, the bulk topological properties do not change, but the
Weyl orbits can be turned off. The joint effect of the Weyl orbits and
surface Lifshitz transition can make the QHE quite complicated. We
find that when the Weyl orbits are broken by the surface Dirac cone, the bulk
and surface states can form the Landau levels (LLs) and contribute the QHE,
independently. The resulting Hall plateaus are sensitive to the thickness of
the sample, but their widths and quantized values are irregularly distributed.
The rest of this paper is organized as follows. In Sec. \ref{MH}, we introduce
the model Hamiltonian and bulk spectrum. The Lifshitz transition of the Fermi
arcs and the LLs are analyzed in Sec. \ref{EFH} and Sec. \ref{ELLs},
respectively. The QHE is studied in Sec. \ref{HEF} and the last section
contains a short summary.

\section{Hamiltonian and bulk spectrum}

\label{MH}

\begin{figure}[ptb]
\centering
%Requires \usepackage{graphicx}
\includegraphics[width=\linewidth]{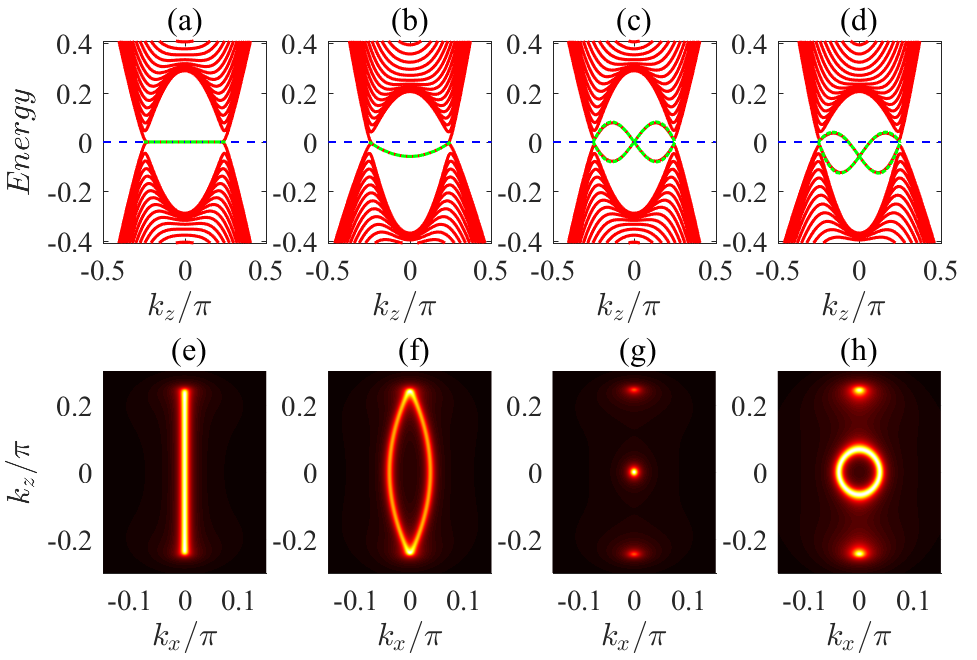}\newline\caption{ (a)-(d) The
dispersion (red) by diagonalizing Hamiltonian (\ref{eq_Hy}) for (a) $\alpha
=0$, $c_{1}=c_{2}=0$, (b) $\alpha=0$, $c_{1}=3c_{2}=0.15$, (c) $\alpha=0.5$,
$c_{1}=c_{2}=0$, and (d) $\alpha=0.5$, $c_{1}=3c_{2}=0.15$, with $k_{x}=0$ and
$E_{D}=0$ denoted by the blue-dashed lines. The green dotted curves are
determined from Eq. (\ref{eq_SFS}), self-consistently. (e)-(h) The
$\boldsymbol{k}$-resolved DOSs corresponding to (a)-(d) at $E_{F}=E_{D}$. The
rest parameters are set as $\lambda=0.5$, $k_{w}=\pi/4$, $m_{0}=\pi^{2}/32$,
$m_{2}=0.5$ and $N_{y}=50$.}%
\label{fig_DS}%
\end{figure}

We begin with a low-energy effective Hamiltonian for the topological DSMs%
\begin{equation}
\mathcal{H}(\boldsymbol{k})=\varepsilon_{\boldsymbol{k}}+\lambda(k_{x}%
\sigma_{z}\tau_{x}-k_{y}\tau_{y})+m_{\boldsymbol{k}}\tau_{z}+\Lambda
(\boldsymbol{k}) \label{eq_Heff}%
\end{equation}
with $m_{\boldsymbol{k}}=m_{0}-m_{1}k_{z}^{2}-m_{2}k_{\parallel}^{2}$ and
$\varepsilon_{\boldsymbol{k}}=c_{0}+c_{1}k_{z}^{2}+c_{2}k_{\parallel}^{2}$,
where $k_{\parallel}=\sqrt{k_{x}^{2}+k_{y}^{2}}$ and $\sigma_{x,y,z}$
($\tau_{x,y,z}$) is the Pauli matrix acting on the spin (orbital parity)
degree of freedom. This model has been widely-adopted to capture the
topological properties of topological DSMs \textrm{Cd}$_{\mathrm{3}}%
$\textrm{As}$_{\mathrm{2}}$\cite{Cd3As2_PhysRevB.88.125427} and \textrm{A}%
$_{\mathrm{3}}$\textrm{Bi} (\textrm{A} = \textrm{Na}, \textrm{K},
\textrm{Rb})\cite{A3B_PhysRevB.85.195320}. In the absence of $\Lambda
(\boldsymbol{k})$, $\left[  \sigma_{z},\mathcal{H}(\boldsymbol{k})\right]  =0$
and the topological DSMs characterized by Hamiltonian (\ref{eq_Heff}) can be
viewed as two superposed copies of a Weyl semimetal with two Weyl nodes, which
possesses two sets of surface Fermi arcs in the surface Brillouin zone, as
illustrated by Figs. \ref{fig_DS}(a)-(b) and (e)-(f). $\Lambda(\boldsymbol{k}%
)$ mixes the eigenstates of opposite spins away from the Dirac points and
plays the role of a momentum-dependent gap term, whose form is determined by
the crystal symmetries.

Specifically, for a DSM with fourfold rotational symmetry, such as
\textrm{Cd}$_{\mathrm{3}}$\textrm{As}$_{\mathrm{2}}$, the momentum-dependent
gap term can take the form $\Lambda(\boldsymbol{k})=\alpha k_{z}\left(
k_{+}^{2}\sigma_{-}+k_{-}^{2}\sigma_{+}\right)  \tau_{x}/2$ with $k_{\pm
}=k_{x}\pm ik_{y}$ and $\sigma_{\pm}=\sigma_{x}\pm i\sigma_{y}$. Diagonalizing
Hamiltonian (\ref{eq_Heff}) yields the continuum bulk spectrum%
\begin{equation}
E_{\pm}(\boldsymbol{k})=\varepsilon_{\boldsymbol{k}}\pm\sqrt{\lambda
^{2}k_{\parallel}^{2}+m_{\boldsymbol{k}}^{2}+\alpha^{2}k_{\parallel}^{4}%
k_{z}^{2}}, \label{eq_BS}%
\end{equation}
from which we can determine the energy location $E_{D}=c_{0}+c_{1}k_{w}^{2}$
and momentum locations $\boldsymbol{K}_{\pm}=(0,0,\pm k_{w})$ of the bulk
Dirac points with $k_{w}=\sqrt{m_{0}/m_{1}}$. As $\varepsilon_{\boldsymbol{k}%
}$ possesses the symmetries of $m_{\boldsymbol{k}}$, it does not qualitatively
change the Dirac spectrum in the bulk, but introduces an asymmetry between the
positive (electrons) and negative (holes) energy branches and, consequently,
breaks the particle-hole symmetry. The electron-hole asymmetry will curve the
Fermi arcs, which was demonstrated to be crucial for the LLs around the Weyl
nodes\cite{PhysRevLett.119.136806}. While $\Lambda(\boldsymbol{k})$ can
profoundly change the spectrum of quasiparticles for sufficiently large
$k_{\parallel}$, it preserves all the symmetries of the crystal structure and
vanishes at the Dirac points. Therefore, the bulk Dirac points are robust
against $\Lambda(\boldsymbol{k})$, as seen from Eq. (\ref{eq_BS}) that both
the momentum and energy locations of the Dirac points are regardless of
$\alpha$. For this reason, $\Lambda(\boldsymbol{k})$ was usually treated as a
bulk perturbation, which does not destroy the bulk topology.

For the states close to the bulk Dirac points, the quasiparticles can be
described by linearizing Hamiltonian (\ref{eq_Heff}) around $\boldsymbol{K}%
_{\pm}$, such that $\Lambda(\boldsymbol{k})$ can be neglected. However, for a
topological DSM slab, the Fermi arcs, which connect the bulk Dirac points
separated far away in momentum space, can extend to large $\boldsymbol{k}$
where the spectrum can be dramatically modified by the momentum-dependent gap
term. In the following, we elucidate that, in response to the
momentum-dependent gap term, the surface states of topological DSMs will
experience a Lifshitz transition, during which process, the Fermi arcs can
exist without connecting the bulk Dirac points.

\begin{figure}[ptb]
\centering
%Requires \usepackage{graphicx}
\includegraphics[width=\linewidth]{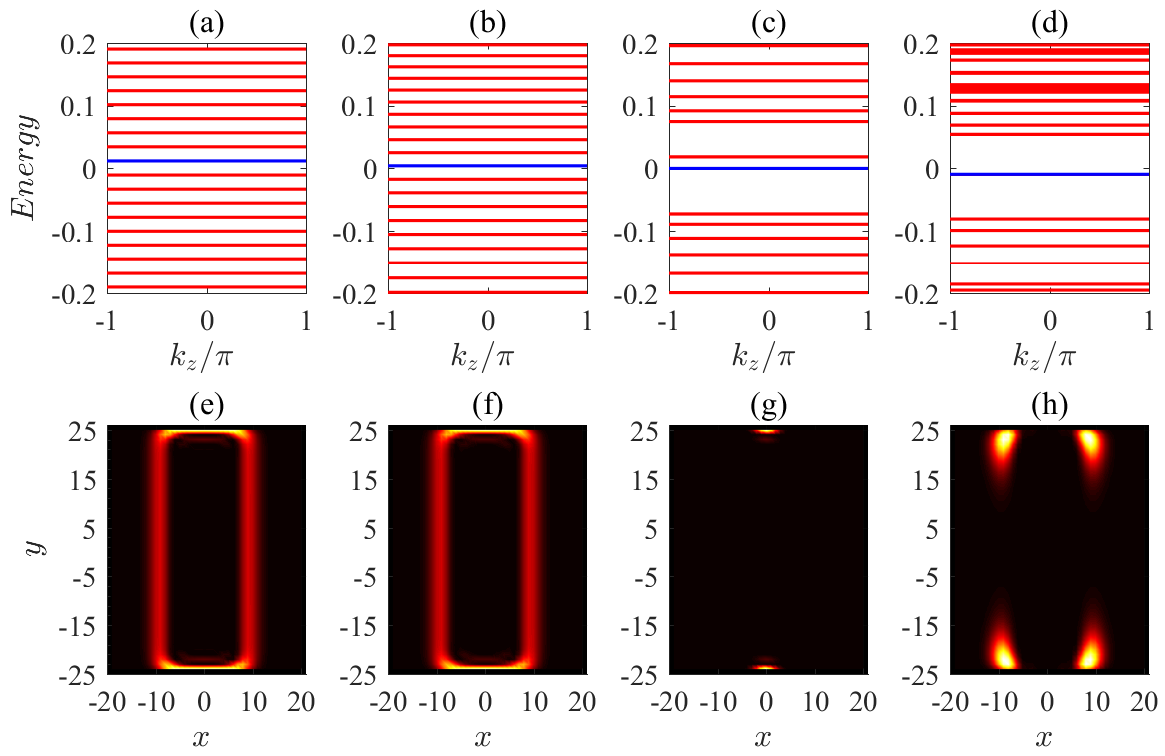}\newline\caption{The LLs for (a)
$\alpha=0$, $c_{1}=c_{2}=0$, (b) $\alpha=0$, $c_{1}=3c_{2}=0.15$, (c)
$\alpha=0.5$, $c_{1}=c_{2}=0$, and (d) $\alpha=0.5$, $c_{1}=3c_{2}=0.15$.
(e)-(h) The probability density $|\psi_{k_{z},\mu}|^{2}$ for the blue LLs in
(a)-(d), in which $\psi_{k_{z},\mu}$ is the $\mu$-th LL's wavefunction. Here,
we set $2\pi\ell_{B}^{2}=N_{x}$, $k_{z}=\pi$, $k_{w}=\pi/2$, $N_{x}=40$ and
other parameters the same as Fig. \ref{fig_DS}.}%
\label{fig_LP}%
\end{figure}

\section{Lifshitz transition of the Fermi arcs}

\label{EFH}

To better understand how the Fermi arc surface states evolve with the
momentum-dependent gap term, we consider a topological DSM slab with open
boundaries at $y=\pm L/2$ and derive the surface states for the $x$-$z$ plane.
For such a finite-size system, $k_{y}$ in Eq. (\ref{eq_Heff}) is not longer a
good quantum number and should be replaced with the operator $-i\partial_{y}$.
In the spirit of the perturbation theory, we construct an unperturbed surface
basis by the surface wavefunctions without $\Lambda(\boldsymbol{k})$ and the
effective Hamiltonian for the surface states can be obtained by projecting Eq.
(\ref{eq_Heff}) onto the unperturbed surface basis. For the convenience of
discussion, we assume $m_{2}>|c_{2}|$ and select $E_{D}$ as potential energy
zero, i.e., $c_{0}=-c_{1}k_{w}$.

By setting $\alpha=0$ and solving the differential equation $\mathcal{H}%
$$(k_{x},-i\partial_{y},k_{z})\Psi_{k_{x},k_{z}}(y)=E\Psi_{k_{x},k_{z}}(y)$
with the open boundary conditions $\Psi_{k_{x},k_{z}}(\pm L/2)=0$, we can
evaluate the unperturbed surface wavefunctions as $\Psi_{k_{x},k_{z}}%
^{\beta,s}(y)=\left(  \frac{1+s}{2},\frac{1-s}{2}\right)  ^{T}\otimes
\Psi_{\beta}(y)$, which are spin-resolved, with $s=\pm1$ being the eigenvalue
of $\sigma_{z}$ and
\begin{equation}
\Psi_{\beta}(y)=\left(
\begin{array}
[c]{c}%
\beta\cos\frac{\vartheta}{2}\\
-\sin\frac{\vartheta}{2}%
\end{array}
\right)  \frac{e^{\kappa_{-}(\beta y-\frac{L}{2})}-e^{\kappa_{+}(\beta
y-\frac{L}{2})}}{\sqrt{N_{\beta}}}. \label{eq_SBS}%
\end{equation}
Here, $\beta=\pm$\emph{ }corresponds to the surface at $y=\pm L/2$, $N_{\beta
}=\int_{-\frac{L}{2}}^{\frac{L}{2}}|e^{\kappa_{-}(\beta y-\frac{L}{2}%
)}-e^{\kappa_{+}(\beta y-\frac{L}{2})}|^{2}dy$ is the normalization
coefficient and%
\begin{equation}
\vartheta=\tan^{-1}\left[  \frac{2\lambda\left(  m_{2}-c_{2}\right)  \left(
\kappa_{+}+\kappa_{-}\right)  }{\lambda^{2}-\left(  m_{2}-c_{2}\right)
^{2}\left(  \kappa_{+}+\kappa_{-}\right)  ^{2}}\right]  ,
\end{equation}
in which $\kappa_{\pm}$ is a solution to $E_{\pm}(k_{x},-i\kappa,k_{z})=E$,
reading as
\begin{equation}
\kappa_{\pm}=\sqrt{k_{x}^{2}+\frac{\zeta_{\lambda}\pm\left(  \zeta_{\lambda
}^{2}-\zeta_{+}\zeta_{-}\right)  ^{1/2}}{m_{2}^{2}-c_{2}^{2}}}
\label{eq_Kappa}%
\end{equation}
with $\zeta_{\lambda}=\frac{\lambda^{2}-\zeta_{+}+\zeta_{-}}{2}$ and%
\begin{equation}
\zeta_{\pm}=\left(  m_{2}\pm c_{2}\right)  \left[  \left(  m_{1}\mp
c_{1}\right)  \left(  k_{w}^{2}-k_{z}^{2}\right)  \mp E\right]  .
\label{eq_Lmc}%
\end{equation}
The surface states are confined within the region defined by
$\operatorname{Re}(\kappa_{\pm})>0$. Subsequently, by performing the
projection operation%
\begin{equation}
\mathcal{H}_{\mathrm{surf}}^{\beta,ss^{\prime}}=\langle\Psi_{k_{x},k_{z}%
}^{\beta,s}(y)|\mathcal{H}(k_{x},-i\partial_{y},k_{z})|\Psi_{k_{x},k_{z}%
}^{\beta,s^{\prime}}(y)\rangle,
\end{equation}
we can obtain the effective surface Hamiltonian%
\begin{equation}
\mathcal{H}_{\mathrm{surf}}^{\beta}\left(  k_{x},k_{z}\right)  =\tilde
{\varepsilon}_{\boldsymbol{k}}-\beta\sin\vartheta\left(  \lambda k_{x}%
\sigma_{z}+\tilde{\alpha}k_{z}\sigma_{x}\right)  , \label{eq_Hsf}%
\end{equation}
where $\tilde{\alpha}=\alpha\left(  k_{x}^{2}-\kappa_{+}\kappa_{-}\right)  $,
$\tilde{\varepsilon}_{\boldsymbol{k}}=\tilde{c}_{1}\left(  k_{z}^{2}-k_{w}%
^{2}\right)  +\tilde{c}_{2}\left(  k_{x}^{2}+\kappa_{+}\kappa_{-}\right)  $
and $\tilde{c}_{l}=c_{l}-m_{l}\cos\vartheta$ with $l=1,2$.

As shown by Eq. (\ref{eq_Hsf}), the surface Hamiltonian exhibits a 2D Dirac
structure with spin-momentum locking, which resembles the surface states for
3D topological insulators. However, as different from the 3D topological
insulators, the bulk spectrum here is also gapless, namely, the surface Dirac
point can coexist with the bulk Dirac points. By diagonalizing Eq.
(\ref{eq_Hsf}), we can arrive at%
\begin{equation}
E_{\eta}^{\beta}\left(  k_{x},k_{z}\right)  =\tilde{\varepsilon}%
_{\boldsymbol{k}}-\eta\beta\sin\vartheta\sqrt{\lambda^{2}k_{x}^{2}%
+\tilde{\alpha}^{2}k_{z}^{2}}, \label{eq_SFS}%
\end{equation}
where $\eta=\pm1$ labels the conduction/valence band. It should be noted that,
because $\kappa_{\pm}$ is energy dependent, Eq. (\ref{eq_SFS}) is only a
formal solution for the surface spectrum. The exact surface dispersion
involves a self-consistent calculation of Eq. (\ref{eq_SFS}) via replacing
$E\rightarrow E_{\eta}^{\beta}\left(  k_{x},k_{z}\right)  $ in Eq.
(\ref{eq_Lmc}). The numerical results of the surface dispersion are presented
by the green dotted curves in Figs. \ref{fig_DS}(a)-(d).

In order to show the bulk states and surface Fermi arcs, simultaneously, we
evaluate the $\boldsymbol{k}$-resolved density of states (DOSs) for a given
energy $E$ through
\begin{equation}
\rho\left(  E,k_{x},k_{z}\right)  =-\frac{1}{\pi}\operatorname{Im}%
\mathrm{Tr}\left(  \frac{1}{E+i0^{+}-H}\right)  ,
\end{equation}
in which $H=\sum_{\boldsymbol{k}}c_{\boldsymbol{k}}^{\dag}\mathcal{H}%
_{\boldsymbol{k}}^{\mathrm{tb}}c_{\boldsymbol{k}}$ is the tight-binding
Hamiltonian corresponding to Eq. (\ref{eq_Heff}). Here, $c_{\boldsymbol{k}%
}^{\dag}$ ($c_{\boldsymbol{k}}$) is the fermion creation (annihilation)
operator and%
\begin{align}
\mathcal{H}_{\boldsymbol{k}}^{\mathrm{tb}} &  =M_{\boldsymbol{k}}%
+\lambda\left(  \sin k_{x}\sigma_{z}\tau_{x}-\sin k_{y}\tau_{y}\right)
-2\alpha\sin k_{z}\nonumber\\
&  \times\left[  \left(  \cos k_{x}-\cos k_{y}\right)  \sigma_{x}-\sin
k_{x}\sin k_{y}\sigma_{y}\right]  \tau_{x}\label{eq_Htb}%
\end{align}
is the single-particle Hamiltonian obtained from Eq. (\ref{eq_Heff}) via the
transforms $k_{a}\rightarrow\sin k_{a}$ and $k_{a}^{2}\rightarrow2-2\cos
k_{a}$, with $a=x,y,z$, $M_{\boldsymbol{k}}=f_{1}\left(  \cos k_{z}-\cos
k_{w}\right)  +f_{2}\left(  \cos k_{x}+\cos k_{y}-2\right)  $ and
$f_{l}=2(m_{l}\tau_{z}-c_{l})$. By performing the Fourier transform
$c_{\boldsymbol{k}}\rightarrow\sum_{n}c_{n}e^{-ik_{y}y_{n}}$, we can
discretize the tight-binding Hamiltonian as%
\begin{equation}
H=\sum_{n}c_{n}^{\dag}h_{0}c_{n}+\sum_{n}\left(  c_{n}^{\dag}h_{y}%
c_{n+1}+h.c.\right)  ,\label{eq_Hy}%
\end{equation}
where the hopping matrices are given by%
\begin{align}
h_{0} &  =f_{0}+f_{1}\cos k_{z}+f_{2}\cos k_{x}+f_{x}\sin k_{x}\nonumber\\
&  -\alpha\sigma_{x}\tau_{x}\left[  \sin\left(  k_{z}+k_{x}\right)
+\sin\left(  k_{z}-k_{x}\right)  \right]  \\
h_{y} &  =\frac{f_{2}+if_{y}}{2}+\alpha\sigma_{x}\tau_{x}\sin k_{z}\nonumber\\
&  +i\alpha\sigma_{y}\tau_{x}\frac{\cos\left(  k_{z}+k_{x}\right)
-\cos\left(  k_{z}-k_{x}\right)  }{2}%
\end{align}
with $f_{0}=-2f_{2}-f_{1}\cos k_{w}$ and $f_{x(y)}=\lambda\sigma_{z(0)}%
\tau_{x(y)}$.

In Fig. \ref{fig_DS}, we plot the numerical spectrum and $\boldsymbol{k}%
$-resolved DOSs for Eq. (\ref{eq_Hy}). As shown by the green dotted curves in
Figs. \ref{fig_DS}(a)-(d), the self-consistent calculation of Eq.
(\ref{eq_SFS}) coincides with the results by the numerical diagonalization of
Eq. (\ref{eq_Hy}). As illustrated by Figs. \ref{fig_DS}(a) and (e), the
surface spectrums without $\varepsilon_{\boldsymbol{k}}$ and $\Lambda
(\boldsymbol{k})$ are $k_{z}$-independent and cross at $E_{D}$ with a
quadruply degenerate line connecting the two Dirac points. The flat Fermi line
can be easily modified by perturbations. This explains why the particle-hole
symmetry and the momentum-dependent gap term are important to the surface
states. A nonzero $\varepsilon_{\boldsymbol{k}}$ can not gap the surface
spectrum, but will bend the Fermi lines to a parabola, as shown by Fig.
\ref{fig_DS}(b). As a result, the Fermi arcs with opposite spin, due to the
spin-dependent term in Eq. (\ref{eq_Hsf}), will curve in opposite direction at
the Fermi level, forming a closed loop with a discontinuous kink at the Dirac
points, as indicated by Fig. \ref{fig_DS}(f).

By contrast, when the momentum-dependent gap term is included, the Fermi lines
with opposite spin, because of the noncommutation between $\sigma_{z}$ and
$\mathcal{H}_{\mathrm{surf}}^{\beta}\left(  k_{x},k_{z}\right)  $, will repel
each other and so remove the spin degeneracy. Since $\Lambda(\boldsymbol{k}%
)=0$ for $k_{\parallel}=0$ or $k_{z}=0$, the surface spectrums keep crossing
at $\boldsymbol{k}=0,\boldsymbol{K}_{\pm}$, as shown by Fig. \ref{fig_DS}(c).
Therefore, a 2D Dirac point develops on each surface, which coexists with the
bulk Dirac points. The surface Dirac points, similar to the bulk Dirac points,
are robust against $\Lambda(\boldsymbol{k})$, but, differently, the energy
location of the surface Dirac points can be modulated by the particle-hole
asymmetry. In the presence of particle-hole symmetry, i.e., $\varepsilon
_{\boldsymbol{k}}=0$, the energy location of the surface Dirac points is
identical to the bulk Dirac points, such that at $E_{F}=E_{D}$, the Fermi line
in Fig. \ref{fig_DS}(e) deforms into three Fermi points in Fig. \ref{fig_DS}%
(g). When the particle-hole symmetry is broken by a finite $\varepsilon
_{\boldsymbol{k}}$, the surface Dirac points will shift away from $E_{D}$,
after that the surface Fermi point around $\boldsymbol{k}=0$ turns to a Fermi
loop, as demonstrated by Fig. \ref{fig_DS}(h), which prevents the Fermi arcs
from connecting the bulk Dirac points.

Consequently, although the surface spectrums remain intersecting with the bulk
Dirac points, the surface states will experience a Lifshitz transition with
the Fermi arcs changing from a curve to some discrete gap-closing points and
further to a Fermi loop coexisting with the bulk Dirac points, during which
process the Weyl orbits can be turned off without destroying the topological
properties of the bulk energy band. \begin{figure}[ptb]
\centering
%Requires \usepackage{graphicx}
\includegraphics[width=\linewidth]{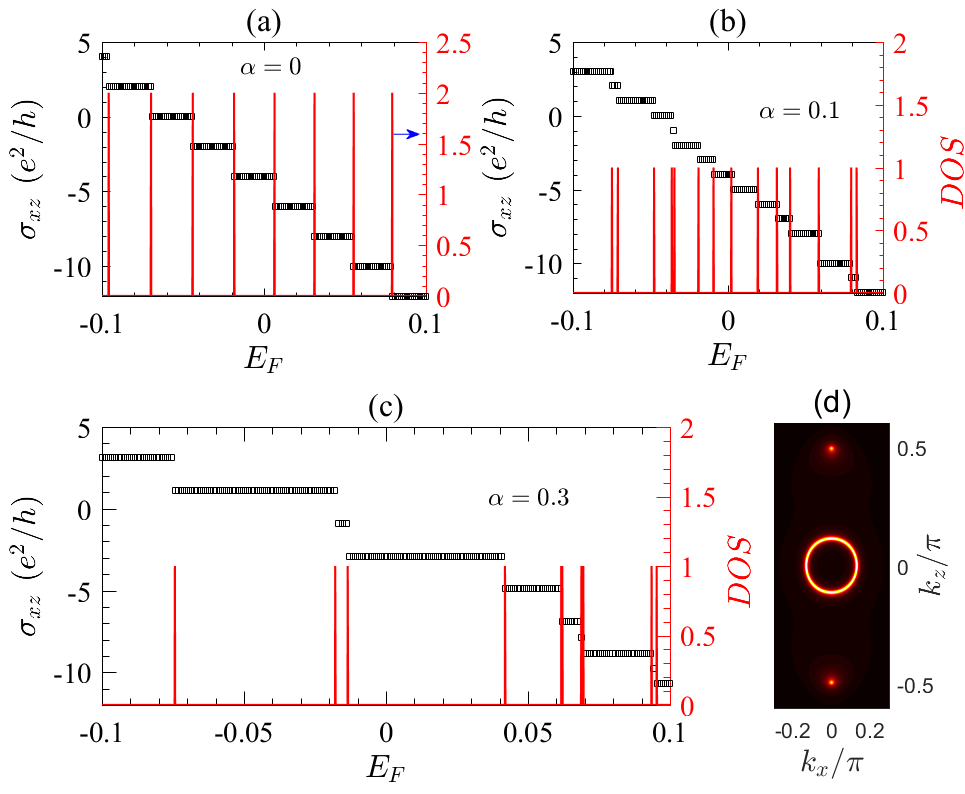}\newline\caption{The Hall
conductivity (left axis) and DOS (right axis) for a given $k_{z}$ with (a)
$\alpha=0$, (b) $\alpha=0.1$ and (c) $\alpha=0.3$. (d) The Fermi surface for
$E_{F}=0$ and $\alpha=0.3$ in the absence of magnetic field. The parameters
are the same as Fig. \ref{fig_LP}.}%
\label{fig_HC}%
\end{figure}

\section{LLs and probability density distribution}

\label{ELLs}

Upon application of an external magnetic field, a Peierls phase should be
added to the hopping matrices as electrons jump from position $\boldsymbol{r}%
_{n}$ to $\boldsymbol{r}_{m}$, i.e., $t_{mn}\rightarrow t_{mn}e^{-i\varphi
_{mn}}$ with $\varphi_{mn}=\frac{2\pi}{\Phi_{0}}\int_{\boldsymbol{r}_{m}%
}^{\boldsymbol{r}_{n}}\boldsymbol{A}\cdot d\boldsymbol{l}$ and $\Phi_{0}=h/e$
being the unit flux quantum. For the magnetic field is applied in the $y$
direction, we can choose the Landau gauge $\boldsymbol{A}=(0,0,-Bx)$. To
include the magnetic field, we further discretize Eq. (\ref{eq_Hy}) in
direction $x$ as%
\begin{align}
H  &  =\sum_{mn}c_{m,n}^{\dag}(T_{m}c_{m,n}+T_{m,x}c_{m+1,n}+T_{m,y}%
c_{m,n+1}\nonumber\\
&  +T_{m,\alpha}c_{m+1,n-1}-T_{m,\alpha}c_{m+1,n+1})+h.c., \label{eq_Hxy}%
\end{align}
in which the Fourier transform $c_{n}\rightarrow\sum_{m}c_{m,n}e^{-ik_{x}%
x_{m}}$ has been adopted and the hopping matrices are%
\begin{align}
T_{m}  &  =\frac{f_{0}+f_{1}\cos\left(  k_{z}-m/\ell_{B}^{2}\right)  }{2},\\
T_{m,\alpha}  &  =\frac{\alpha}{2}\sigma_{y}\tau_{x}\sin\left(  k_{z}%
-\frac{m+1/2}{\ell_{B}^{2}}\right)  ,\label{eq_Tx}\\
T_{m,y}  &  =\frac{f_{2}+if_{y}}{2}+\alpha\sigma_{x}\tau_{x}\sin\left(
k_{z}-m/\ell_{B}^{2}\right)  ,\\
T_{m,x}  &  =\frac{f_{2}-if_{x}}{2}-\alpha\sigma_{y}\tau_{x}\sin\left(
k_{z}-\frac{m+1/2}{\ell_{B}^{2}}\right)  , \label{eq_Tpm}%
\end{align}
with $\ell_{B}=\sqrt{\hbar/\left(  eB\right)  }$ being the magnetic length.

After introducing the magnetic field, we continue to demonstrate the relation
between Eq. (\ref{eq_Htb}) and Eq. (\ref{eq_Hxy}). By the inverse Fourier
transform on Eq. (\ref{eq_Hxy}), we can arrive at%
\begin{align}
\mathcal{\tilde{H}}_{\boldsymbol{k}} &  =f_{1}\left(  \cos\kappa_{z}-\cos
k_{w}\right)  +f_{2}\left(  \cos k_{x}+\cos k_{y}-2\right)  \nonumber\\
&  +\lambda\left(  \sin k_{x}\tau_{x}\sigma_{z}-\sin k_{y}\tau_{y}\right)
-\alpha\tau_{x}[(\frac{\sin\kappa_{z}^{+}-\sin\kappa_{z}^{-}}{2}\nonumber\\
&  -\cos k_{y}\sin\kappa_{z})\sigma_{x}-\sin k_{y}\frac{\cos\kappa_{z}%
^{+}-\cos\kappa_{z}^{-}}{2}\sigma_{y}]\label{eq_Heffk}%
\end{align}
with $\kappa_{z}=k_{z}-x/\ell_{B}^{2}$ and $\kappa_{z}^{\pm}=\pm\kappa
_{z}+k_{x}$. Here, we have used the Baker-Campbell-Hausdorff formula, i.e.,
$e^{\hat{A}+\hat{B}}=e^{\hat{A}}e^{\hat{B}}e^{-\frac{1}{2}[\hat{A},\hat{B}]}$.
It is clear that $\mathcal{\tilde{H}}_{\boldsymbol{k}}=\mathcal{H}%
_{\boldsymbol{k}\rightarrow\boldsymbol{k}+e\boldsymbol{A}/\hbar}^{\mathrm{tb}%
}$ follows the Peierls substitution. However, for $\alpha\neq0$, one should be
very careful when using the direct Peierls substitution, because of the cross
momentum term in $\Lambda(\boldsymbol{k})$. For example, as we transform Eq.
(\ref{eq_Heffk}) to the lattice form for numerical calculation, there will
emerge an additional phase $\ell_{B}^{-2}/2$, such as in Eqs. (\ref{eq_Tx})
and (\ref{eq_Tpm}). The underlying physics for the additional phase is that
after the Peierls substitution, the different momentum components can be
noncommutative, e.g., $[k_{x},\kappa_{z}]=$ $i\ell_{B}^{-2}$, and the
Baker-Campbell-Hausdorff formula must be adopted. For instance, when
transforming Eq. (\ref{eq_Hy}) to Eq. (\ref{eq_Hxy}), one should express the
trigonometric functions in the exponential form and then introduce the Peierls
phase via the transforms $e^{ik_{z}}\rightarrow e^{i\kappa_{z}}$ and
$e^{i\left(  k_{x}\pm k_{z}\right)  }\rightarrow e^{i\kappa_{z}^{\pm}}=e^{\pm
i\left(  \kappa_{z}-\ell_{B}^{-2}/2\right)  }e^{ik_{x}}$.

By diagonalizing Hamiltonian (\ref{eq_Hxy}) numerically, we can obtain the LLs
and spatial distribution of the electron probability density, as displayed in
Fig. \ref{fig_LP}. Through the spatial distribution of the probability
density, we can tell easily whether or not the LLs are formed by the Weyl
orbits. From Figs. \ref{fig_LP}(a) and (e), we see that the LLs can be formed
from the Weyl-orbit mechanism, even when there is no curvature in the Fermi
arcs. In the Weyl-orbit mechanism, the surface fermions, driven by the
magnetic field, will move along the Fermi arcs from one Dirac valley to the
other and tunnel to the opposite surface at the Dirac points via the bulk
states. Therefore, the probability density exhibits a closed loop with two
bright stripes crossing the bulk and connecting the surface states, as seen
from Fig. \ref{fig_LP}(e). The width of the bright stripes $\sim2\ell_{B}$
relates to the cyclotron radius of the bulk Dirac fermions and the distance
between the stripes encodes the momentum distance between the Dirac points.
The cyclotron center $x_{c}=\ell_{B}^{2}k_{z}$ of the fermions in different
Dirac valleys differs by $\Delta x_{c}=2\ell_{B}^{2}k_{w}$, which is exactly
the distance between the two bright stripes.

\begin{figure}[ptb]
\centering
%Requires \usepackage{graphicx}
\includegraphics[width=\linewidth]{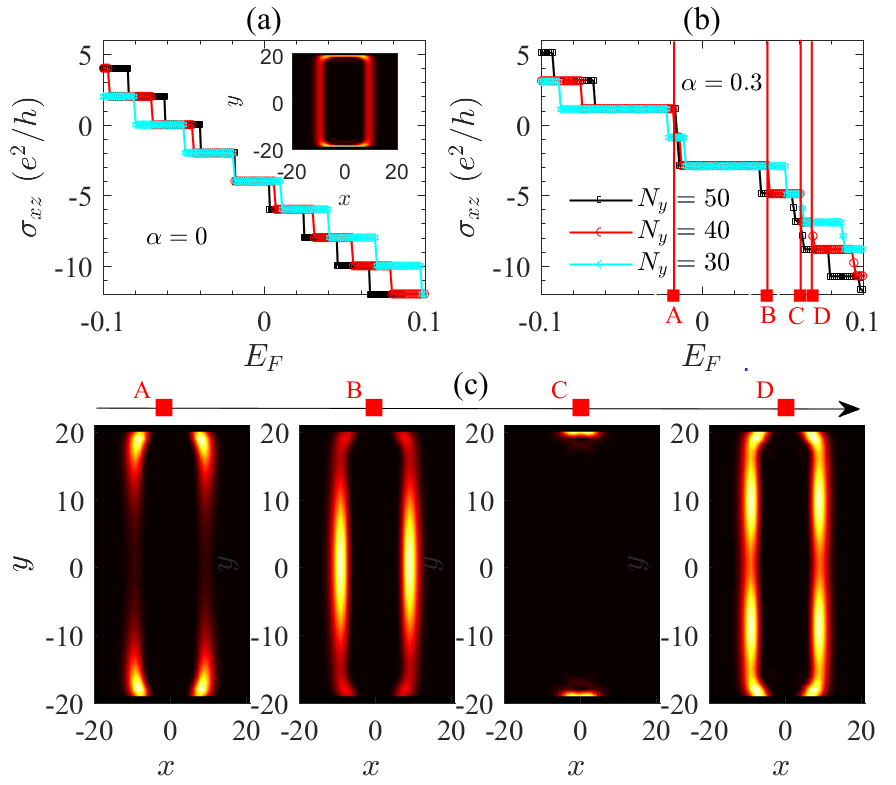}\newline\caption{(a)-(b)
Evolution of the Hall conductivity with the sample thickness, i.e.,
$N_{y}=(30,40,50)$, for (a) $\alpha=0$ with the inset showing that the LLs are
from the Weyl-orbit mechanism, and (b) $\alpha=0.3$ with the red lines marked
by $A$-$D$ denoting the LLs for $N_{y}=40$. (c) The probability density of the
LLs marked in (b), which indicates that the surface and bulk states are
disconnected, i.e., the Weyl orbits have been turned off. The parameters are
chosen the same as Fig. \ref{fig_HC}. }%
\label{fig_PP}%
\end{figure}

A finite $\varepsilon_{\boldsymbol{k}}$ that curves the Fermi arcs in the
surface Brillouin zone will shift the LLs, integrally, by modifying the
magnetic flux enclosed by the Weyl orbits, as demonstrated by Figs.
\ref{fig_LP}(b) and (f). In this case, the Weyl orbits will not be destroyed,
so that the LLs remain evenly spaced. As expected, the LLs will respond
dramatically to $\Lambda(\boldsymbol{k})$, see Figs. \ref{fig_LP}(c) and (d)
where the LLs distribute irregularly, for the Weyl-orbit mechanism can be
turned off by the surface Dirac cone. The probability densities displayed in
Figs. \ref{fig_LP}(g) and (h), as well as Fig. \ref{fig_PP}(c), illustrate
that the bulk and surface states can support the cyclotron orbit and form the
LLs, independently. The superposition of the bulk and surface LLs results in
the complicated LLs in Figs. \ref{fig_LP}(c) and (d). 

\section{Evolution of the QHE with the surface Lifshitz transition}

\label{HEF}

As discussed above, in the presence of the momentum-dependent gap term, the
Weyl orbit near the Dirac points will vanish. However, the LLs remain
discrete, implying that the QHE in topological DSMs can happen without
involving the Weyl-orbit mechanism. As $k_{z}$ is a good quantum number, we
can calculate the Hall conductivity from the Kubo
formula\cite{PhysRevLett.119.136806,PhysRevLett.125.036602,PhysRevB.104.075425,PhysRevB.104.205305,PhysRevB.103.245434}%
\begin{align}
\sigma_{xz} &  =\frac{i\hbar e^{2}}{L_{x}L_{z}}\sum_{k_{z},\mu\neq\nu}\left(
f_{\nu,k_{z}}-f_{\mu,k_{z}}\right)  \nonumber\\
&  \times\frac{\langle\psi_{k_{z},\mu}|\hat{\upsilon}_{x}|\psi_{k_{z},\nu
}\rangle\langle\psi_{k_{z},\nu}|\hat{\upsilon}_{z}|\psi_{k_{z},\mu}\rangle
}{\left(  \varepsilon_{k_{z},\mu}-\varepsilon_{k_{z},\nu}\right)  ^{2}%
},\label{eq_Kubo}%
\end{align}
where $\varepsilon_{\mu,k_{z}}$ denote respectively the $\mu$-th energy of Eq.
(\ref{eq_Hxy}), and the velocity operators are defined as $\hat{\upsilon}%
_{x}=i\hbar^{-1}\sum_{mn}\left[  H,c_{m,n}^{\dag}x_{m}c_{m,n}\right]  $ and
$\hat{\upsilon}_{z}=\partial H/(\hbar\partial k_{z})$. Here, $f_{\mu,k_{z}%
}=[1+\exp(\frac{\varepsilon_{\mu,k_{z}}-E_{F}}{k_{B}T})]^{-1}$ is the
Fermi-Dirac distribution function, with $k_{B}$ and $T$ being the Boltzmann's
constant and temperature, respectively.

In Fig. \ref{fig_HC}, we present the numerical results for the Hall
conductivity near the bulk Dirac points. As can be seen, the Hall conductivity
exhibits a step-wise structure and jumps whenever the Fermi level goes across
a LL. The Hall plateaus for $\alpha=0$ are quantized as $2ne^{2}/h$, with $n$
as an integer and the factor $2$ owing to the spin degeneracy of the LLs. The
spin degeneracy of the LLs can be characterized by the $k_{z}$-resolved DOS
plotted in the right axis of Fig. \ref{fig_HC}. In the absence of the
momentum-dependent gap term, the LLs are evenly spaced and therefore, the
widths of the Hall plateaus are identical, as shown by Fig. \ref{fig_HC}(a).
With increasing $\alpha$, the Weyl orbits near the bulk Dirac points will be
broken gradually, and as a result, the LLs become complicated and the widths
of the Hall plateaus turn to be irregular, as observed in Figs. \ref{fig_HC}%
(b) and (c). With the Weyl orbits destroyed, the spin degeneracy of the LLs
will be removed, leading to the emergence of odd Hall plateaus in Figs.
\ref{fig_HC}(b) and (c).

The thickness dependence of the quantized Hall plateaus is one of the
characteristic signals for the 3D QHE \cite{zhang2019quantum}. The
Weyl-orbit-mediated 3D QHE is thickness-dependent, as shown by Fig.
\ref{fig_PP}(a), where the Hall plateaus increase synchronously in width as
$N_{y}$ decreases. However, from Fig. \ref{fig_PP}(b), we see that when the
Weyl orbits have been turned off by the surface Dirac cone, the Hall plateaus
can still be sensitive to the thickness of the sample. The thickness-dependent
QHE in the absence of the Weyl orbits can be attributable to the bulk LLs, as
a result of the energy quantization by the confinement in direction $y$, as
illustrated by Fig. \ref{fig_PP}(c), where the probability density exhibits a
standing wave configuration in analogous to the one for the infinite well. In
contrast to the bulk LLs, the surface LLs are less sensitive to the thickness,
as shown by the LL marked by $C$ in Fig. \ref{fig_PP}(b). While the surface
LLs are thickness-independent, the Hall plateau close to them can display
thickness dependence for the surface LLs are surrounded by the bulk LLs.

\section{Conclusion}

In summary, we have investigated the QHE in topological DSMs under
consideration of the surface Lifshitz transition modulated by a bulk
momentum-dependent gap term. It is found that the bulk momentum-dependent gap
term, as a higher-order momentum term, can be neglected in study of the bulk
topological properties, but it will dramatically deform the Fermi arcs and
lead to the surface Lifshitz transition. During the surface Lifshitz
transition, a 2D Dirac cone will develop for the surface states, which keeps
the Fermi arcs from connecting the bulk Dirac points and, as a result, the
Weyl orbits can be turned off without breaking the topology of the bulk energy
band. In response to the Lifshitz transition, the 3D QHE because of the Weyl
orbits will break down, along with that the quantized Hall plateaus turn to be
irregular. As a bulk perturbation, the momentum-dependent gap term is quite
common in real materials. Therefore, in addition to the bulk Weyl nodes and
surface Fermi arcs, the Lifshitz transition of the surface states also plays
an important part in realization of stable Weyl orbits.

\section{acknowledgements}

This work was supported by the National NSF of China under Grants No.
12274146, No. 11874016, No. 12174121 and No. 12104167, the Guangdong Basic and
Applied Basic Research Foundation under Grant No. 2023B1515020050, the Science
and Technology Program of Guangzhou under Grant No. 2019050001 and the
Guangdong Provincial Key Laboratory under Grant No. 2020B1212060066.

\bibliographystyle{apsrev4-1}
\bibliography{bibQTR}

\begin{thebibliography}{54}%
\makeatletter
\providecommand \@ifxundefined [1]{%
 \@ifx{#1\undefined}
}%
\providecommand \@ifnum [1]{%
 \ifnum #1\expandafter \@firstoftwo
 \else \expandafter \@secondoftwo
 \fi
}%
\providecommand \@ifx [1]{%
 \ifx #1\expandafter \@firstoftwo
 \else \expandafter \@secondoftwo
 \fi
}%
\providecommand \natexlab [1]{#1}%
\providecommand \enquote  [1]{``#1''}%
\providecommand \bibnamefont  [1]{#1}%
\providecommand \bibfnamefont [1]{#1}%
\providecommand \citenamefont [1]{#1}%
\providecommand \href@noop [0]{\@secondoftwo}%
\providecommand \href [0]{\begingroup \@sanitize@url \@href}%
\providecommand \@href[1]{\@@startlink{#1}\@@href}%
\providecommand \@@href[1]{\endgroup#1\@@endlink}%
\providecommand \@sanitize@url [0]{\catcode `\\12\catcode `\$12\catcode
  `\&12\catcode `\#12\catcode `\^12\catcode `\_12\catcode `\%12\relax}%
\providecommand \@@startlink[1]{}%
\providecommand \@@endlink[0]{}%
\providecommand \url  [0]{\begingroup\@sanitize@url \@url }%
\providecommand \@url [1]{\endgroup\@href {#1}{\urlprefix }}%
\providecommand \urlprefix  [0]{URL }%
\providecommand \Eprint [0]{\href }%
\providecommand \doibase [0]{http://dx.doi.org/}%
\providecommand \selectlanguage [0]{\@gobble}%
\providecommand \bibinfo  [0]{\@secondoftwo}%
\providecommand \bibfield  [0]{\@secondoftwo}%
\providecommand \translation [1]{[#1]}%
\providecommand \BibitemOpen [0]{}%
\providecommand \bibitemStop [0]{}%
\providecommand \bibitemNoStop [0]{.\EOS\space}%
\providecommand \EOS [0]{\spacefactor3000\relax}%
\providecommand \BibitemShut  [1]{\csname bibitem#1\endcsname}%
\let\auto@bib@innerbib\@empty
%</preamble>
\bibitem [{\citenamefont {Armitage}\ \emph {et~al.}(2018)\citenamefont
  {Armitage}, \citenamefont {Mele},\ and\ \citenamefont
  {Vishwanath}}]{RevModPhys.90.015001}%
  \BibitemOpen
  \bibfield  {author} {\bibinfo {author} {\bibfnamefont {N.~P.}\ \bibnamefont
  {Armitage}}, \bibinfo {author} {\bibfnamefont {E.~J.}\ \bibnamefont {Mele}},
  \ and\ \bibinfo {author} {\bibfnamefont {A.}~\bibnamefont {Vishwanath}},\
  }\href {\doibase 10.1103/RevModPhys.90.015001} {\bibfield  {journal}
  {\bibinfo  {journal} {Rev. Mod. Phys.}\ }\textbf {\bibinfo {volume} {90}},\
  \bibinfo {pages} {015001} (\bibinfo {year} {2018})}\BibitemShut {NoStop}%
\bibitem [{\citenamefont {Neupane}\ \emph {et~al.}(2014)\citenamefont
  {Neupane}, \citenamefont {Xu}, \citenamefont {Sankar}, \citenamefont
  {Alidoust}, \citenamefont {Bian}, \citenamefont {Liu}, \citenamefont
  {Belopolski}, \citenamefont {Chang}, \citenamefont {Jeng}, \citenamefont
  {Lin}, \citenamefont {Bansil}, \citenamefont {Chou},\ and\ \citenamefont
  {Hasan}}]{Neupane:2014aa}%
  \BibitemOpen
  \bibfield  {author} {\bibinfo {author} {\bibfnamefont {M.}~\bibnamefont
  {Neupane}}, \bibinfo {author} {\bibfnamefont {S.-Y.}\ \bibnamefont {Xu}},
  \bibinfo {author} {\bibfnamefont {R.}~\bibnamefont {Sankar}}, \bibinfo
  {author} {\bibfnamefont {N.}~\bibnamefont {Alidoust}}, \bibinfo {author}
  {\bibfnamefont {G.}~\bibnamefont {Bian}}, \bibinfo {author} {\bibfnamefont
  {C.}~\bibnamefont {Liu}}, \bibinfo {author} {\bibfnamefont {I.}~\bibnamefont
  {Belopolski}}, \bibinfo {author} {\bibfnamefont {T.-R.}\ \bibnamefont
  {Chang}}, \bibinfo {author} {\bibfnamefont {H.-T.}\ \bibnamefont {Jeng}},
  \bibinfo {author} {\bibfnamefont {H.}~\bibnamefont {Lin}}, \bibinfo {author}
  {\bibfnamefont {A.}~\bibnamefont {Bansil}}, \bibinfo {author} {\bibfnamefont
  {F.}~\bibnamefont {Chou}}, \ and\ \bibinfo {author} {\bibfnamefont {M.~Z.}\
  \bibnamefont {Hasan}},\ }\href {https://doi.org/10.1038/ncomms4786}
  {\bibfield  {journal} {\bibinfo  {journal} {Nat. Commun.}\ }\textbf {\bibinfo
  {volume} {5}},\ \bibinfo {pages} {3786} (\bibinfo {year} {2014})}\BibitemShut
  {NoStop}%
\bibitem [{\citenamefont {Wang}\ \emph {et~al.}(2013)\citenamefont {Wang},
  \citenamefont {Weng}, \citenamefont {Wu}, \citenamefont {Dai},\ and\
  \citenamefont {Fang}}]{Cd3As2_PhysRevB.88.125427}%
  \BibitemOpen
  \bibfield  {author} {\bibinfo {author} {\bibfnamefont {Z.}~\bibnamefont
  {Wang}}, \bibinfo {author} {\bibfnamefont {H.}~\bibnamefont {Weng}}, \bibinfo
  {author} {\bibfnamefont {Q.}~\bibnamefont {Wu}}, \bibinfo {author}
  {\bibfnamefont {X.}~\bibnamefont {Dai}}, \ and\ \bibinfo {author}
  {\bibfnamefont {Z.}~\bibnamefont {Fang}},\ }\href {\doibase
  10.1103/PhysRevB.88.125427} {\bibfield  {journal} {\bibinfo  {journal} {Phys.
  Rev. B}\ }\textbf {\bibinfo {volume} {88}},\ \bibinfo {pages} {125427}
  (\bibinfo {year} {2013})}\BibitemShut {NoStop}%
\bibitem [{\citenamefont {Wang}\ \emph {et~al.}(2012)\citenamefont {Wang},
  \citenamefont {Sun}, \citenamefont {Chen}, \citenamefont {Franchini},
  \citenamefont {Xu}, \citenamefont {Weng}, \citenamefont {Dai},\ and\
  \citenamefont {Fang}}]{A3B_PhysRevB.85.195320}%
  \BibitemOpen
  \bibfield  {author} {\bibinfo {author} {\bibfnamefont {Z.}~\bibnamefont
  {Wang}}, \bibinfo {author} {\bibfnamefont {Y.}~\bibnamefont {Sun}}, \bibinfo
  {author} {\bibfnamefont {X.-Q.}\ \bibnamefont {Chen}}, \bibinfo {author}
  {\bibfnamefont {C.}~\bibnamefont {Franchini}}, \bibinfo {author}
  {\bibfnamefont {G.}~\bibnamefont {Xu}}, \bibinfo {author} {\bibfnamefont
  {H.}~\bibnamefont {Weng}}, \bibinfo {author} {\bibfnamefont {X.}~\bibnamefont
  {Dai}}, \ and\ \bibinfo {author} {\bibfnamefont {Z.}~\bibnamefont {Fang}},\
  }\href {\doibase 10.1103/PhysRevB.85.195320} {\bibfield  {journal} {\bibinfo
  {journal} {Phys. Rev. B}\ }\textbf {\bibinfo {volume} {85}},\ \bibinfo
  {pages} {195320} (\bibinfo {year} {2012})}\BibitemShut {NoStop}%
\bibitem [{\citenamefont {Liu}\ \emph {et~al.}(2014)\citenamefont {Liu},
  \citenamefont {Zhou}, \citenamefont {Zhang}, \citenamefont {Wang},
  \citenamefont {Weng}, \citenamefont {Prabhakaran}, \citenamefont {Mo},
  \citenamefont {Shen}, \citenamefont {Fang}, \citenamefont {Dai},
  \citenamefont {Hussain},\ and\ \citenamefont {Chen}}]{Liu864}%
  \BibitemOpen
  \bibfield  {author} {\bibinfo {author} {\bibfnamefont {Z.~K.}\ \bibnamefont
  {Liu}}, \bibinfo {author} {\bibfnamefont {B.}~\bibnamefont {Zhou}}, \bibinfo
  {author} {\bibfnamefont {Y.}~\bibnamefont {Zhang}}, \bibinfo {author}
  {\bibfnamefont {Z.~J.}\ \bibnamefont {Wang}}, \bibinfo {author}
  {\bibfnamefont {H.~M.}\ \bibnamefont {Weng}}, \bibinfo {author}
  {\bibfnamefont {D.}~\bibnamefont {Prabhakaran}}, \bibinfo {author}
  {\bibfnamefont {S.-K.}\ \bibnamefont {Mo}}, \bibinfo {author} {\bibfnamefont
  {Z.~X.}\ \bibnamefont {Shen}}, \bibinfo {author} {\bibfnamefont
  {Z.}~\bibnamefont {Fang}}, \bibinfo {author} {\bibfnamefont {X.}~\bibnamefont
  {Dai}}, \bibinfo {author} {\bibfnamefont {Z.}~\bibnamefont {Hussain}}, \ and\
  \bibinfo {author} {\bibfnamefont {Y.~L.}\ \bibnamefont {Chen}},\ }\href
  {\doibase 10.1126/science.1245085} {\bibfield  {journal} {\bibinfo  {journal}
  {Science}\ }\textbf {\bibinfo {volume} {343}},\ \bibinfo {pages} {864}
  (\bibinfo {year} {2014})}\BibitemShut {NoStop}%
\bibitem [{\citenamefont {Yang}\ and\ \citenamefont
  {Nagaosa}(2014)}]{Yang:2014aa}%
  \BibitemOpen
  \bibfield  {author} {\bibinfo {author} {\bibfnamefont {B.-J.}\ \bibnamefont
  {Yang}}\ and\ \bibinfo {author} {\bibfnamefont {N.}~\bibnamefont {Nagaosa}},\
  }\href {\doibase 10.1038/ncomms5898} {\bibfield  {journal} {\bibinfo
  {journal} {Nat. Commun.}\ }\textbf {\bibinfo {volume} {5}},\ \bibinfo {pages}
  {4898} (\bibinfo {year} {2014})}\BibitemShut {NoStop}%
\bibitem [{\citenamefont {Wan}\ \emph {et~al.}(2011)\citenamefont {Wan},
  \citenamefont {Turner}, \citenamefont {Vishwanath},\ and\ \citenamefont
  {Savrasov}}]{PhysRevB.83.205101}%
  \BibitemOpen
  \bibfield  {author} {\bibinfo {author} {\bibfnamefont {X.}~\bibnamefont
  {Wan}}, \bibinfo {author} {\bibfnamefont {A.~M.}\ \bibnamefont {Turner}},
  \bibinfo {author} {\bibfnamefont {A.}~\bibnamefont {Vishwanath}}, \ and\
  \bibinfo {author} {\bibfnamefont {S.~Y.}\ \bibnamefont {Savrasov}},\ }\href
  {\doibase 10.1103/PhysRevB.83.205101} {\bibfield  {journal} {\bibinfo
  {journal} {Phys. Rev. B}\ }\textbf {\bibinfo {volume} {83}},\ \bibinfo
  {pages} {205101} (\bibinfo {year} {2011})}\BibitemShut {NoStop}%
\bibitem [{\citenamefont {Zhang}\ \emph {et~al.}(2016)\citenamefont {Zhang},
  \citenamefont {Xu}, \citenamefont {Belopolski}, \citenamefont {Yuan},
  \citenamefont {Lin}, \citenamefont {Tong}, \citenamefont {Bian},
  \citenamefont {Alidoust}, \citenamefont {Lee}, \citenamefont {Huang},
  \citenamefont {Chang}, \citenamefont {Chang}, \citenamefont {Hsu},
  \citenamefont {Jeng}, \citenamefont {Neupane}, \citenamefont {Sanchez},
  \citenamefont {Zheng}, \citenamefont {Wang}, \citenamefont {Lin},
  \citenamefont {Zhang}, \citenamefont {Lu}, \citenamefont {Shen},
  \citenamefont {Neupert}, \citenamefont {Zahid~Hasan},\ and\ \citenamefont
  {Jia}}]{Zhang:2016aa}%
  \BibitemOpen
  \bibfield  {author} {\bibinfo {author} {\bibfnamefont {C.-L.}\ \bibnamefont
  {Zhang}}, \bibinfo {author} {\bibfnamefont {S.-Y.}\ \bibnamefont {Xu}},
  \bibinfo {author} {\bibfnamefont {I.}~\bibnamefont {Belopolski}}, \bibinfo
  {author} {\bibfnamefont {Z.}~\bibnamefont {Yuan}}, \bibinfo {author}
  {\bibfnamefont {Z.}~\bibnamefont {Lin}}, \bibinfo {author} {\bibfnamefont
  {B.}~\bibnamefont {Tong}}, \bibinfo {author} {\bibfnamefont {G.}~\bibnamefont
  {Bian}}, \bibinfo {author} {\bibfnamefont {N.}~\bibnamefont {Alidoust}},
  \bibinfo {author} {\bibfnamefont {C.-C.}\ \bibnamefont {Lee}}, \bibinfo
  {author} {\bibfnamefont {S.-M.}\ \bibnamefont {Huang}}, \bibinfo {author}
  {\bibfnamefont {T.-R.}\ \bibnamefont {Chang}}, \bibinfo {author}
  {\bibfnamefont {G.}~\bibnamefont {Chang}}, \bibinfo {author} {\bibfnamefont
  {C.-H.}\ \bibnamefont {Hsu}}, \bibinfo {author} {\bibfnamefont {H.-T.}\
  \bibnamefont {Jeng}}, \bibinfo {author} {\bibfnamefont {M.}~\bibnamefont
  {Neupane}}, \bibinfo {author} {\bibfnamefont {D.~S.}\ \bibnamefont
  {Sanchez}}, \bibinfo {author} {\bibfnamefont {H.}~\bibnamefont {Zheng}},
  \bibinfo {author} {\bibfnamefont {J.}~\bibnamefont {Wang}}, \bibinfo {author}
  {\bibfnamefont {H.}~\bibnamefont {Lin}}, \bibinfo {author} {\bibfnamefont
  {C.}~\bibnamefont {Zhang}}, \bibinfo {author} {\bibfnamefont {H.-Z.}\
  \bibnamefont {Lu}}, \bibinfo {author} {\bibfnamefont {S.-Q.}\ \bibnamefont
  {Shen}}, \bibinfo {author} {\bibfnamefont {T.}~\bibnamefont {Neupert}},
  \bibinfo {author} {\bibfnamefont {M.}~\bibnamefont {Zahid~Hasan}}, \ and\
  \bibinfo {author} {\bibfnamefont {S.}~\bibnamefont {Jia}},\ }\href
  {https://doi.org/10.1038/ncomms10735} {\bibfield  {journal} {\bibinfo
  {journal} {Nat. Commun.}\ }\textbf {\bibinfo {volume} {7}},\ \bibinfo {pages}
  {10735} (\bibinfo {year} {2016})}\BibitemShut {NoStop}%
\bibitem [{\citenamefont {Zhang}\ \emph {et~al.}(2017)\citenamefont {Zhang},
  \citenamefont {Zhang}, \citenamefont {Wang}, \citenamefont {Liu},
  \citenamefont {Chen}, \citenamefont {Lu}, \citenamefont {Liang},
  \citenamefont {Cao}, \citenamefont {Yuan}, \citenamefont {Tang},
  \citenamefont {Li}, \citenamefont {Zhou}, \citenamefont {Gu}, \citenamefont
  {Wu}, \citenamefont {Zou},\ and\ \citenamefont {Xiu}}]{Zhang:2017aa}%
  \BibitemOpen
  \bibfield  {author} {\bibinfo {author} {\bibfnamefont {C.}~\bibnamefont
  {Zhang}}, \bibinfo {author} {\bibfnamefont {E.}~\bibnamefont {Zhang}},
  \bibinfo {author} {\bibfnamefont {W.}~\bibnamefont {Wang}}, \bibinfo {author}
  {\bibfnamefont {Y.}~\bibnamefont {Liu}}, \bibinfo {author} {\bibfnamefont
  {Z.-G.}\ \bibnamefont {Chen}}, \bibinfo {author} {\bibfnamefont
  {S.}~\bibnamefont {Lu}}, \bibinfo {author} {\bibfnamefont {S.}~\bibnamefont
  {Liang}}, \bibinfo {author} {\bibfnamefont {J.}~\bibnamefont {Cao}}, \bibinfo
  {author} {\bibfnamefont {X.}~\bibnamefont {Yuan}}, \bibinfo {author}
  {\bibfnamefont {L.}~\bibnamefont {Tang}}, \bibinfo {author} {\bibfnamefont
  {Q.}~\bibnamefont {Li}}, \bibinfo {author} {\bibfnamefont {C.}~\bibnamefont
  {Zhou}}, \bibinfo {author} {\bibfnamefont {T.}~\bibnamefont {Gu}}, \bibinfo
  {author} {\bibfnamefont {Y.}~\bibnamefont {Wu}}, \bibinfo {author}
  {\bibfnamefont {J.}~\bibnamefont {Zou}}, \ and\ \bibinfo {author}
  {\bibfnamefont {F.}~\bibnamefont {Xiu}},\ }\href
  {https://doi.org/10.1038/ncomms13741} {\bibfield  {journal} {\bibinfo
  {journal} {Nat. Commun.}\ }\textbf {\bibinfo {volume} {8}},\ \bibinfo {pages}
  {13741} (\bibinfo {year} {2017})}\BibitemShut {NoStop}%
\bibitem [{\citenamefont {Wang}\ \emph {et~al.}(2017)\citenamefont {Wang},
  \citenamefont {Sun}, \citenamefont {Lu},\ and\ \citenamefont
  {Xie}}]{PhysRevLett.119.136806}%
  \BibitemOpen
  \bibfield  {author} {\bibinfo {author} {\bibfnamefont {C.~M.}\ \bibnamefont
  {Wang}}, \bibinfo {author} {\bibfnamefont {H.-P.}\ \bibnamefont {Sun}},
  \bibinfo {author} {\bibfnamefont {H.-Z.}\ \bibnamefont {Lu}}, \ and\ \bibinfo
  {author} {\bibfnamefont {X.~C.}\ \bibnamefont {Xie}},\ }\href {\doibase
  10.1103/PhysRevLett.119.136806} {\bibfield  {journal} {\bibinfo  {journal}
  {Phys. Rev. Lett.}\ }\textbf {\bibinfo {volume} {119}},\ \bibinfo {pages}
  {136806} (\bibinfo {year} {2017})}\BibitemShut {NoStop}%
\bibitem [{\citenamefont {Gorbar}\ \emph {et~al.}(2015)\citenamefont {Gorbar},
  \citenamefont {Miransky}, \citenamefont {Shovkovy},\ and\ \citenamefont
  {Sukhachov}}]{PhysRevB.91.121101}%
  \BibitemOpen
  \bibfield  {author} {\bibinfo {author} {\bibfnamefont {E.~V.}\ \bibnamefont
  {Gorbar}}, \bibinfo {author} {\bibfnamefont {V.~A.}\ \bibnamefont
  {Miransky}}, \bibinfo {author} {\bibfnamefont {I.~A.}\ \bibnamefont
  {Shovkovy}}, \ and\ \bibinfo {author} {\bibfnamefont {P.~O.}\ \bibnamefont
  {Sukhachov}},\ }\href {\doibase 10.1103/PhysRevB.91.121101} {\bibfield
  {journal} {\bibinfo  {journal} {Phys. Rev. B}\ }\textbf {\bibinfo {volume}
  {91}},\ \bibinfo {pages} {121101} (\bibinfo {year} {2015})}\BibitemShut
  {NoStop}%
\bibitem [{\citenamefont {Kargarian}\ \emph {et~al.}(2016)\citenamefont
  {Kargarian}, \citenamefont {Randeria},\ and\ \citenamefont
  {Lu}}]{Kargarian8648}%
  \BibitemOpen
  \bibfield  {author} {\bibinfo {author} {\bibfnamefont {M.}~\bibnamefont
  {Kargarian}}, \bibinfo {author} {\bibfnamefont {M.}~\bibnamefont {Randeria}},
  \ and\ \bibinfo {author} {\bibfnamefont {Y.-M.}\ \bibnamefont {Lu}},\ }\href
  {\doibase 10.1073/pnas.1524787113} {\bibfield  {journal} {\bibinfo  {journal}
  {Proc. Natl. Acad. Sci. USA}\ }\textbf {\bibinfo {volume} {113}},\ \bibinfo
  {pages} {8648} (\bibinfo {year} {2016})}\BibitemShut {NoStop}%
\bibitem [{\citenamefont {Nielsen}\ and\ \citenamefont
  {Ninomiya}(1983)}]{NIELSEN1983389}%
  \BibitemOpen
  \bibfield  {author} {\bibinfo {author} {\bibfnamefont {H.}~\bibnamefont
  {Nielsen}}\ and\ \bibinfo {author} {\bibfnamefont {M.}~\bibnamefont
  {Ninomiya}},\ }\href {\doibase https://doi.org/10.1016/0370-2693(83)91529-0}
  {\bibfield  {journal} {\bibinfo  {journal} {Physics Letters B}\ }\textbf
  {\bibinfo {volume} {130}},\ \bibinfo {pages} {389 } (\bibinfo {year}
  {1983})}\BibitemShut {NoStop}%
\bibitem [{\citenamefont {Volovik}(2003)}]{Volovik2003}%
  \BibitemOpen
  \bibfield  {author} {\bibinfo {author} {\bibfnamefont {G.~E.}\ \bibnamefont
  {Volovik}},\ }\href@noop {} {\emph {\bibinfo {title} {The universe in a
  helium droplet}}},\ Vol.\ \bibinfo {volume} {117}\ (\bibinfo  {publisher}
  {Oxford University Press on Demand},\ \bibinfo {year} {2003})\BibitemShut
  {NoStop}%
\bibitem [{\citenamefont {Raza}\ \emph {et~al.}(2019)\citenamefont {Raza},
  \citenamefont {Sirota},\ and\ \citenamefont {Teo}}]{PhysRevX.9.011039}%
  \BibitemOpen
  \bibfield  {author} {\bibinfo {author} {\bibfnamefont {S.}~\bibnamefont
  {Raza}}, \bibinfo {author} {\bibfnamefont {A.}~\bibnamefont {Sirota}}, \ and\
  \bibinfo {author} {\bibfnamefont {J.~C.~Y.}\ \bibnamefont {Teo}},\ }\href
  {\doibase 10.1103/PhysRevX.9.011039} {\bibfield  {journal} {\bibinfo
  {journal} {Phys. Rev. X}\ }\textbf {\bibinfo {volume} {9}},\ \bibinfo {pages}
  {011039} (\bibinfo {year} {2019})}\BibitemShut {NoStop}%
\bibitem [{\citenamefont {Zyuzin}\ \emph {et~al.}(2012)\citenamefont {Zyuzin},
  \citenamefont {Wu},\ and\ \citenamefont {Burkov}}]{PhysRevB.85.165110}%
  \BibitemOpen
  \bibfield  {author} {\bibinfo {author} {\bibfnamefont {A.~A.}\ \bibnamefont
  {Zyuzin}}, \bibinfo {author} {\bibfnamefont {S.}~\bibnamefont {Wu}}, \ and\
  \bibinfo {author} {\bibfnamefont {A.~A.}\ \bibnamefont {Burkov}},\ }\href
  {\doibase 10.1103/PhysRevB.85.165110} {\bibfield  {journal} {\bibinfo
  {journal} {Phys. Rev. B}\ }\textbf {\bibinfo {volume} {85}},\ \bibinfo
  {pages} {165110} (\bibinfo {year} {2012})}\BibitemShut {NoStop}%
\bibitem [{\citenamefont {Goswami}\ and\ \citenamefont
  {Tewari}(2013)}]{PhysRevB.88.245107}%
  \BibitemOpen
  \bibfield  {author} {\bibinfo {author} {\bibfnamefont {P.}~\bibnamefont
  {Goswami}}\ and\ \bibinfo {author} {\bibfnamefont {S.}~\bibnamefont
  {Tewari}},\ }\href {\doibase 10.1103/PhysRevB.88.245107} {\bibfield
  {journal} {\bibinfo  {journal} {Phys. Rev. B}\ }\textbf {\bibinfo {volume}
  {88}},\ \bibinfo {pages} {245107} (\bibinfo {year} {2013})}\BibitemShut
  {NoStop}%
\bibitem [{\citenamefont {Han}\ \emph {et~al.}(2018)\citenamefont {Han},
  \citenamefont {Cho},\ and\ \citenamefont {Moon}}]{PhysRevB.98.085149}%
  \BibitemOpen
  \bibfield  {author} {\bibinfo {author} {\bibfnamefont {S.}~\bibnamefont
  {Han}}, \bibinfo {author} {\bibfnamefont {G.~Y.}\ \bibnamefont {Cho}}, \ and\
  \bibinfo {author} {\bibfnamefont {E.-G.}\ \bibnamefont {Moon}},\ }\href
  {\doibase 10.1103/PhysRevB.98.085149} {\bibfield  {journal} {\bibinfo
  {journal} {Phys. Rev. B}\ }\textbf {\bibinfo {volume} {98}},\ \bibinfo
  {pages} {085149} (\bibinfo {year} {2018})}\BibitemShut {NoStop}%
\bibitem [{\citenamefont {Deng}\ \emph {et~al.}(2017)\citenamefont {Deng},
  \citenamefont {Luo}, \citenamefont {Wang}, \citenamefont {Sheng},\ and\
  \citenamefont {Xing}}]{PhysRevB.96.155141}%
  \BibitemOpen
  \bibfield  {author} {\bibinfo {author} {\bibfnamefont {M.-X.}\ \bibnamefont
  {Deng}}, \bibinfo {author} {\bibfnamefont {W.}~\bibnamefont {Luo}}, \bibinfo
  {author} {\bibfnamefont {R.-Q.}\ \bibnamefont {Wang}}, \bibinfo {author}
  {\bibfnamefont {L.}~\bibnamefont {Sheng}}, \ and\ \bibinfo {author}
  {\bibfnamefont {D.~Y.}\ \bibnamefont {Xing}},\ }\href {\doibase
  10.1103/PhysRevB.96.155141} {\bibfield  {journal} {\bibinfo  {journal} {Phys.
  Rev. B}\ }\textbf {\bibinfo {volume} {96}},\ \bibinfo {pages} {155141}
  (\bibinfo {year} {2017})}\BibitemShut {NoStop}%
\bibitem [{\citenamefont {Chen}\ \emph {et~al.}(2019)\citenamefont {Chen},
  \citenamefont {Yu}, \citenamefont {Li}, \citenamefont {Chen}, \citenamefont
  {Sheng},\ and\ \citenamefont {Yang}}]{PhysRevB.99.075131}%
  \BibitemOpen
  \bibfield  {author} {\bibinfo {author} {\bibfnamefont {C.}~\bibnamefont
  {Chen}}, \bibinfo {author} {\bibfnamefont {Z.-M.}\ \bibnamefont {Yu}},
  \bibinfo {author} {\bibfnamefont {S.}~\bibnamefont {Li}}, \bibinfo {author}
  {\bibfnamefont {Z.}~\bibnamefont {Chen}}, \bibinfo {author} {\bibfnamefont
  {X.-L.}\ \bibnamefont {Sheng}}, \ and\ \bibinfo {author} {\bibfnamefont
  {S.~A.}\ \bibnamefont {Yang}},\ }\href {\doibase 10.1103/PhysRevB.99.075131}
  {\bibfield  {journal} {\bibinfo  {journal} {Phys. Rev. B}\ }\textbf {\bibinfo
  {volume} {99}},\ \bibinfo {pages} {075131} (\bibinfo {year}
  {2019})}\BibitemShut {NoStop}%
\bibitem [{\citenamefont {Ali}\ \emph {et~al.}(2014)\citenamefont {Ali},
  \citenamefont {Xiong}, \citenamefont {Flynn}, \citenamefont {Tao},
  \citenamefont {Gibson}, \citenamefont {Schoop}, \citenamefont {Liang},
  \citenamefont {Haldolaarachchige}, \citenamefont {Hirschberger},
  \citenamefont {Ong},\ and\ \citenamefont {Cava}}]{Ali2014}%
  \BibitemOpen
  \bibfield  {author} {\bibinfo {author} {\bibfnamefont {M.~N.}\ \bibnamefont
  {Ali}}, \bibinfo {author} {\bibfnamefont {J.}~\bibnamefont {Xiong}}, \bibinfo
  {author} {\bibfnamefont {S.}~\bibnamefont {Flynn}}, \bibinfo {author}
  {\bibfnamefont {J.}~\bibnamefont {Tao}}, \bibinfo {author} {\bibfnamefont
  {Q.~D.}\ \bibnamefont {Gibson}}, \bibinfo {author} {\bibfnamefont {L.~M.}\
  \bibnamefont {Schoop}}, \bibinfo {author} {\bibfnamefont {T.}~\bibnamefont
  {Liang}}, \bibinfo {author} {\bibfnamefont {N.}~\bibnamefont
  {Haldolaarachchige}}, \bibinfo {author} {\bibfnamefont {M.}~\bibnamefont
  {Hirschberger}}, \bibinfo {author} {\bibfnamefont {N.}~\bibnamefont {Ong}}, \
  and\ \bibinfo {author} {\bibfnamefont {R.}~\bibnamefont {Cava}},\ }\href
  {\doibase 10.1038/nature13763} {\bibfield  {journal} {\bibinfo  {journal}
  {Nature}\ }\textbf {\bibinfo {volume} {514}},\ \bibinfo {pages} {205}
  (\bibinfo {year} {2014})}\BibitemShut {NoStop}%
\bibitem [{\citenamefont {Shekhar}\ \emph {et~al.}(2015)\citenamefont
  {Shekhar}, \citenamefont {Nayak}, \citenamefont {Sun}, \citenamefont
  {Schmidt}, \citenamefont {Nicklas}, \citenamefont {Leermakers}, \citenamefont
  {Zeitler}, \citenamefont {Skourski}, \citenamefont {Wosnitza}, \citenamefont
  {Liu}, \citenamefont {Chen}, \citenamefont {Schnelle}, \citenamefont
  {Borrmann}, \citenamefont {Grin}, \citenamefont {Felser},\ and\ \citenamefont
  {Yan}}]{Shekhar2015}%
  \BibitemOpen
  \bibfield  {author} {\bibinfo {author} {\bibfnamefont {C.}~\bibnamefont
  {Shekhar}}, \bibinfo {author} {\bibfnamefont {A.~K.}\ \bibnamefont {Nayak}},
  \bibinfo {author} {\bibfnamefont {Y.}~\bibnamefont {Sun}}, \bibinfo {author}
  {\bibfnamefont {M.}~\bibnamefont {Schmidt}}, \bibinfo {author} {\bibfnamefont
  {M.}~\bibnamefont {Nicklas}}, \bibinfo {author} {\bibfnamefont
  {I.}~\bibnamefont {Leermakers}}, \bibinfo {author} {\bibfnamefont
  {U.}~\bibnamefont {Zeitler}}, \bibinfo {author} {\bibfnamefont
  {Y.}~\bibnamefont {Skourski}}, \bibinfo {author} {\bibfnamefont
  {J.}~\bibnamefont {Wosnitza}}, \bibinfo {author} {\bibfnamefont
  {Z.}~\bibnamefont {Liu}}, \bibinfo {author} {\bibfnamefont {Y.}~\bibnamefont
  {Chen}}, \bibinfo {author} {\bibfnamefont {W.}~\bibnamefont {Schnelle}},
  \bibinfo {author} {\bibfnamefont {H.}~\bibnamefont {Borrmann}}, \bibinfo
  {author} {\bibfnamefont {Y.}~\bibnamefont {Grin}}, \bibinfo {author}
  {\bibfnamefont {C.}~\bibnamefont {Felser}}, \ and\ \bibinfo {author}
  {\bibfnamefont {B.}~\bibnamefont {Yan}},\ }\href {\doibase 10.1038/nphys3372}
  {\bibfield  {journal} {\bibinfo  {journal} {Nat. Phys.}\ }\textbf {\bibinfo
  {volume} {11}},\ \bibinfo {pages} {645} (\bibinfo {year} {2015})}\BibitemShut
  {NoStop}%
\bibitem [{\citenamefont {Parameswaran}\ \emph {et~al.}(2014)\citenamefont
  {Parameswaran}, \citenamefont {Grover}, \citenamefont {Abanin}, \citenamefont
  {Pesin},\ and\ \citenamefont {Vishwanath}}]{PhysRevX.4.031035}%
  \BibitemOpen
  \bibfield  {author} {\bibinfo {author} {\bibfnamefont {S.~A.}\ \bibnamefont
  {Parameswaran}}, \bibinfo {author} {\bibfnamefont {T.}~\bibnamefont
  {Grover}}, \bibinfo {author} {\bibfnamefont {D.~A.}\ \bibnamefont {Abanin}},
  \bibinfo {author} {\bibfnamefont {D.~A.}\ \bibnamefont {Pesin}}, \ and\
  \bibinfo {author} {\bibfnamefont {A.}~\bibnamefont {Vishwanath}},\ }\href
  {\doibase 10.1103/PhysRevX.4.031035} {\bibfield  {journal} {\bibinfo
  {journal} {Phys. Rev. X}\ }\textbf {\bibinfo {volume} {4}},\ \bibinfo {pages}
  {031035} (\bibinfo {year} {2014})}\BibitemShut {NoStop}%
\bibitem [{\citenamefont {Xiong}\ \emph {et~al.}(2015)\citenamefont {Xiong},
  \citenamefont {Kushwaha}, \citenamefont {Liang}, \citenamefont {Krizan},
  \citenamefont {Hirschberger}, \citenamefont {Wang}, \citenamefont {Cava},\
  and\ \citenamefont {Ong}}]{Xiong413}%
  \BibitemOpen
  \bibfield  {author} {\bibinfo {author} {\bibfnamefont {J.}~\bibnamefont
  {Xiong}}, \bibinfo {author} {\bibfnamefont {S.~K.}\ \bibnamefont {Kushwaha}},
  \bibinfo {author} {\bibfnamefont {T.}~\bibnamefont {Liang}}, \bibinfo
  {author} {\bibfnamefont {J.~W.}\ \bibnamefont {Krizan}}, \bibinfo {author}
  {\bibfnamefont {M.}~\bibnamefont {Hirschberger}}, \bibinfo {author}
  {\bibfnamefont {W.}~\bibnamefont {Wang}}, \bibinfo {author} {\bibfnamefont
  {R.~J.}\ \bibnamefont {Cava}}, \ and\ \bibinfo {author} {\bibfnamefont
  {N.~P.}\ \bibnamefont {Ong}},\ }\href {\doibase 10.1126/science.aac6089}
  {\bibfield  {journal} {\bibinfo  {journal} {Science}\ }\textbf {\bibinfo
  {volume} {350}},\ \bibinfo {pages} {413} (\bibinfo {year}
  {2015})}\BibitemShut {NoStop}%
\bibitem [{\citenamefont {Li}\ \emph {et~al.}(2015)\citenamefont {Li},
  \citenamefont {Wang}, \citenamefont {Liu}, \citenamefont {Wang},
  \citenamefont {Liao},\ and\ \citenamefont {Yu}}]{Li:2015aa}%
  \BibitemOpen
  \bibfield  {author} {\bibinfo {author} {\bibfnamefont {C.-Z.}\ \bibnamefont
  {Li}}, \bibinfo {author} {\bibfnamefont {L.-X.}\ \bibnamefont {Wang}},
  \bibinfo {author} {\bibfnamefont {H.}~\bibnamefont {Liu}}, \bibinfo {author}
  {\bibfnamefont {J.}~\bibnamefont {Wang}}, \bibinfo {author} {\bibfnamefont
  {Z.-M.}\ \bibnamefont {Liao}}, \ and\ \bibinfo {author} {\bibfnamefont
  {D.-P.}\ \bibnamefont {Yu}},\ }\href {https://doi.org/10.1038/ncomms10137}
  {\bibfield  {journal} {\bibinfo  {journal} {Nat. Commun.}\ }\textbf {\bibinfo
  {volume} {6}},\ \bibinfo {pages} {10137} (\bibinfo {year}
  {2015})}\BibitemShut {NoStop}%
\bibitem [{\citenamefont {Deng}\ \emph
  {et~al.}(2019{\natexlab{a}})\citenamefont {Deng}, \citenamefont {Qi},
  \citenamefont {Ma}, \citenamefont {Shen}, \citenamefont {Wang}, \citenamefont
  {Sheng},\ and\ \citenamefont {Xing}}]{PhysRevLett.122.036601}%
  \BibitemOpen
  \bibfield  {author} {\bibinfo {author} {\bibfnamefont {M.-X.}\ \bibnamefont
  {Deng}}, \bibinfo {author} {\bibfnamefont {G.~Y.}\ \bibnamefont {Qi}},
  \bibinfo {author} {\bibfnamefont {R.}~\bibnamefont {Ma}}, \bibinfo {author}
  {\bibfnamefont {R.}~\bibnamefont {Shen}}, \bibinfo {author} {\bibfnamefont
  {R.-Q.}\ \bibnamefont {Wang}}, \bibinfo {author} {\bibfnamefont
  {L.}~\bibnamefont {Sheng}}, \ and\ \bibinfo {author} {\bibfnamefont {D.~Y.}\
  \bibnamefont {Xing}},\ }\href {\doibase 10.1103/PhysRevLett.122.036601}
  {\bibfield  {journal} {\bibinfo  {journal} {Phys. Rev. Lett.}\ }\textbf
  {\bibinfo {volume} {122}},\ \bibinfo {pages} {036601} (\bibinfo {year}
  {2019}{\natexlab{a}})}\BibitemShut {NoStop}%
\bibitem [{\citenamefont {Deng}\ \emph
  {et~al.}(2019{\natexlab{b}})\citenamefont {Deng}, \citenamefont {Duan},
  \citenamefont {Luo}, \citenamefont {Deng}, \citenamefont {Wang},\ and\
  \citenamefont {Sheng}}]{PhysRevB.99.165146}%
  \BibitemOpen
  \bibfield  {author} {\bibinfo {author} {\bibfnamefont {M.-X.}\ \bibnamefont
  {Deng}}, \bibinfo {author} {\bibfnamefont {H.-J.}\ \bibnamefont {Duan}},
  \bibinfo {author} {\bibfnamefont {W.}~\bibnamefont {Luo}}, \bibinfo {author}
  {\bibfnamefont {W.~Y.}\ \bibnamefont {Deng}}, \bibinfo {author}
  {\bibfnamefont {R.-Q.}\ \bibnamefont {Wang}}, \ and\ \bibinfo {author}
  {\bibfnamefont {L.}~\bibnamefont {Sheng}},\ }\href {\doibase
  10.1103/PhysRevB.99.165146} {\bibfield  {journal} {\bibinfo  {journal} {Phys.
  Rev. B}\ }\textbf {\bibinfo {volume} {99}},\ \bibinfo {pages} {165146}
  (\bibinfo {year} {2019}{\natexlab{b}})}\BibitemShut {NoStop}%
\bibitem [{\citenamefont {Potter}\ \emph {et~al.}(2014)\citenamefont {Potter},
  \citenamefont {Kimchi},\ and\ \citenamefont
  {Vishwanath}}]{Nat.Commun.5.5161}%
  \BibitemOpen
  \bibfield  {author} {\bibinfo {author} {\bibfnamefont {A.~C.}\ \bibnamefont
  {Potter}}, \bibinfo {author} {\bibfnamefont {I.}~\bibnamefont {Kimchi}}, \
  and\ \bibinfo {author} {\bibfnamefont {A.}~\bibnamefont {Vishwanath}},\
  }\href {\doibase 10.1038/ncomms6161} {\bibfield  {journal} {\bibinfo
  {journal} {Nat. Commun.}\ }\textbf {\bibinfo {volume} {5}},\ \bibinfo {pages}
  {5161} (\bibinfo {year} {2014})}\BibitemShut {NoStop}%
\bibitem [{\citenamefont {Moll}\ \emph {et~al.}(2016)\citenamefont {Moll},
  \citenamefont {Nair}, \citenamefont {Helm}, \citenamefont {Potter},
  \citenamefont {Kimchi}, \citenamefont {Vishwanath},\ and\ \citenamefont
  {Analytis}}]{Moll2016}%
  \BibitemOpen
  \bibfield  {author} {\bibinfo {author} {\bibfnamefont {P.~J.~W.}\
  \bibnamefont {Moll}}, \bibinfo {author} {\bibfnamefont {N.~L.}\ \bibnamefont
  {Nair}}, \bibinfo {author} {\bibfnamefont {T.}~\bibnamefont {Helm}}, \bibinfo
  {author} {\bibfnamefont {A.~C.}\ \bibnamefont {Potter}}, \bibinfo {author}
  {\bibfnamefont {I.}~\bibnamefont {Kimchi}}, \bibinfo {author} {\bibfnamefont
  {A.}~\bibnamefont {Vishwanath}}, \ and\ \bibinfo {author} {\bibfnamefont
  {J.~G.}\ \bibnamefont {Analytis}},\ }\href {\doibase 10.1038/nature18276}
  {\bibfield  {journal} {\bibinfo  {journal} {Nature}\ }\textbf {\bibinfo
  {volume} {535}},\ \bibinfo {pages} {266} (\bibinfo {year}
  {2016})}\BibitemShut {NoStop}%
\bibitem [{\citenamefont {Lin}\ \emph {et~al.}(2017)\citenamefont {Lin},
  \citenamefont {Wang}, \citenamefont {Wang}, \citenamefont {Li}, \citenamefont
  {Li}, \citenamefont {Yu},\ and\ \citenamefont {Liao}}]{PhysRevB.95.235436}%
  \BibitemOpen
  \bibfield  {author} {\bibinfo {author} {\bibfnamefont {B.-C.}\ \bibnamefont
  {Lin}}, \bibinfo {author} {\bibfnamefont {S.}~\bibnamefont {Wang}}, \bibinfo
  {author} {\bibfnamefont {L.-X.}\ \bibnamefont {Wang}}, \bibinfo {author}
  {\bibfnamefont {C.-Z.}\ \bibnamefont {Li}}, \bibinfo {author} {\bibfnamefont
  {J.-G.}\ \bibnamefont {Li}}, \bibinfo {author} {\bibfnamefont
  {D.}~\bibnamefont {Yu}}, \ and\ \bibinfo {author} {\bibfnamefont {Z.-M.}\
  \bibnamefont {Liao}},\ }\href {\doibase 10.1103/PhysRevB.95.235436}
  {\bibfield  {journal} {\bibinfo  {journal} {Phys. Rev. B}\ }\textbf {\bibinfo
  {volume} {95}},\ \bibinfo {pages} {235436} (\bibinfo {year}
  {2017})}\BibitemShut {NoStop}%
\bibitem [{\citenamefont {Wang}\ \emph {et~al.}(2016)\citenamefont {Wang},
  \citenamefont {Li}, \citenamefont {Yu},\ and\ \citenamefont
  {Liao}}]{Wang20166}%
  \BibitemOpen
  \bibfield  {author} {\bibinfo {author} {\bibfnamefont {L.-X.}\ \bibnamefont
  {Wang}}, \bibinfo {author} {\bibfnamefont {C.-Z.}\ \bibnamefont {Li}},
  \bibinfo {author} {\bibfnamefont {D.-P.}\ \bibnamefont {Yu}}, \ and\ \bibinfo
  {author} {\bibfnamefont {Z.-M.}\ \bibnamefont {Liao}},\ }\href {\doibase
  10.1038/ncomms10769} {\bibfield  {journal} {\bibinfo  {journal} {Nature
  Communications}\ }\textbf {\bibinfo {volume} {7}},\ \bibinfo {pages} {10769}
  (\bibinfo {year} {2016})}\BibitemShut {NoStop}%
\bibitem [{\citenamefont {Wang}\ \emph {et~al.}(2018)\citenamefont {Wang},
  \citenamefont {Lin}, \citenamefont {Zheng}, \citenamefont {Yu},\ and\
  \citenamefont {Liao}}]{PhysRevLett.120.257701}%
  \BibitemOpen
  \bibfield  {author} {\bibinfo {author} {\bibfnamefont {S.}~\bibnamefont
  {Wang}}, \bibinfo {author} {\bibfnamefont {B.-C.}\ \bibnamefont {Lin}},
  \bibinfo {author} {\bibfnamefont {W.-Z.}\ \bibnamefont {Zheng}}, \bibinfo
  {author} {\bibfnamefont {D.}~\bibnamefont {Yu}}, \ and\ \bibinfo {author}
  {\bibfnamefont {Z.-M.}\ \bibnamefont {Liao}},\ }\href {\doibase
  10.1103/PhysRevLett.120.257701} {\bibfield  {journal} {\bibinfo  {journal}
  {Phys. Rev. Lett.}\ }\textbf {\bibinfo {volume} {120}},\ \bibinfo {pages}
  {257701} (\bibinfo {year} {2018})}\BibitemShut {NoStop}%
\bibitem [{\citenamefont {Deng}\ \emph {et~al.}(2022)\citenamefont {Deng},
  \citenamefont {Hu}, \citenamefont {Luo}, \citenamefont {Duan},\ and\
  \citenamefont {Wang}}]{PhysRevB.106.075139}%
  \BibitemOpen
  \bibfield  {author} {\bibinfo {author} {\bibfnamefont {M.-X.}\ \bibnamefont
  {Deng}}, \bibinfo {author} {\bibfnamefont {Y.-C.}\ \bibnamefont {Hu}},
  \bibinfo {author} {\bibfnamefont {W.}~\bibnamefont {Luo}}, \bibinfo {author}
  {\bibfnamefont {H.-J.}\ \bibnamefont {Duan}}, \ and\ \bibinfo {author}
  {\bibfnamefont {R.-Q.}\ \bibnamefont {Wang}},\ }\href {\doibase
  10.1103/PhysRevB.106.075139} {\bibfield  {journal} {\bibinfo  {journal}
  {Phys. Rev. B}\ }\textbf {\bibinfo {volume} {106}},\ \bibinfo {pages}
  {075139} (\bibinfo {year} {2022})}\BibitemShut {NoStop}%
\bibitem [{\citenamefont {Li}\ \emph {et~al.}(2020)\citenamefont {Li},
  \citenamefont {Liu}, \citenamefont {Jiang},\ and\ \citenamefont
  {Xie}}]{PhysRevLett.125.036602}%
  \BibitemOpen
  \bibfield  {author} {\bibinfo {author} {\bibfnamefont {H.}~\bibnamefont
  {Li}}, \bibinfo {author} {\bibfnamefont {H.}~\bibnamefont {Liu}}, \bibinfo
  {author} {\bibfnamefont {H.}~\bibnamefont {Jiang}}, \ and\ \bibinfo {author}
  {\bibfnamefont {X.~C.}\ \bibnamefont {Xie}},\ }\href {\doibase
  10.1103/PhysRevLett.125.036602} {\bibfield  {journal} {\bibinfo  {journal}
  {Phys. Rev. Lett.}\ }\textbf {\bibinfo {volume} {125}},\ \bibinfo {pages}
  {036602} (\bibinfo {year} {2020})}\BibitemShut {NoStop}%
\bibitem [{\citenamefont {Ma}\ \emph {et~al.}(2021)\citenamefont {Ma},
  \citenamefont {Sheng},\ and\ \citenamefont {Sheng}}]{PhysRevB.104.075425}%
  \BibitemOpen
  \bibfield  {author} {\bibinfo {author} {\bibfnamefont {R.}~\bibnamefont
  {Ma}}, \bibinfo {author} {\bibfnamefont {D.~N.}\ \bibnamefont {Sheng}}, \
  and\ \bibinfo {author} {\bibfnamefont {L.}~\bibnamefont {Sheng}},\ }\href
  {\doibase 10.1103/PhysRevB.104.075425} {\bibfield  {journal} {\bibinfo
  {journal} {Phys. Rev. B}\ }\textbf {\bibinfo {volume} {104}},\ \bibinfo
  {pages} {075425} (\bibinfo {year} {2021})}\BibitemShut {NoStop}%
\bibitem [{\citenamefont {Geng}\ \emph {et~al.}(2021)\citenamefont {Geng},
  \citenamefont {Qi}, \citenamefont {Sheng}, \citenamefont {Chen},\ and\
  \citenamefont {Xing}}]{PhysRevB.104.205305}%
  \BibitemOpen
  \bibfield  {author} {\bibinfo {author} {\bibfnamefont {H.}~\bibnamefont
  {Geng}}, \bibinfo {author} {\bibfnamefont {G.~Y.}\ \bibnamefont {Qi}},
  \bibinfo {author} {\bibfnamefont {L.}~\bibnamefont {Sheng}}, \bibinfo
  {author} {\bibfnamefont {W.}~\bibnamefont {Chen}}, \ and\ \bibinfo {author}
  {\bibfnamefont {D.~Y.}\ \bibnamefont {Xing}},\ }\href {\doibase
  10.1103/PhysRevB.104.205305} {\bibfield  {journal} {\bibinfo  {journal}
  {Phys. Rev. B}\ }\textbf {\bibinfo {volume} {104}},\ \bibinfo {pages}
  {205305} (\bibinfo {year} {2021})}\BibitemShut {NoStop}%
\bibitem [{\citenamefont {Chang}\ \emph {et~al.}(2021)\citenamefont {Chang},
  \citenamefont {Geng}, \citenamefont {Sheng},\ and\ \citenamefont
  {Xing}}]{PhysRevB.103.245434}%
  \BibitemOpen
  \bibfield  {author} {\bibinfo {author} {\bibfnamefont {M.}~\bibnamefont
  {Chang}}, \bibinfo {author} {\bibfnamefont {H.}~\bibnamefont {Geng}},
  \bibinfo {author} {\bibfnamefont {L.}~\bibnamefont {Sheng}}, \ and\ \bibinfo
  {author} {\bibfnamefont {D.~Y.}\ \bibnamefont {Xing}},\ }\href {\doibase
  10.1103/PhysRevB.103.245434} {\bibfield  {journal} {\bibinfo  {journal}
  {Phys. Rev. B}\ }\textbf {\bibinfo {volume} {103}},\ \bibinfo {pages}
  {245434} (\bibinfo {year} {2021})}\BibitemShut {NoStop}%
\bibitem [{\citenamefont {Chang}\ and\ \citenamefont
  {Sheng}(2021)}]{PhysRevB.103.245409}%
  \BibitemOpen
  \bibfield  {author} {\bibinfo {author} {\bibfnamefont {M.}~\bibnamefont
  {Chang}}\ and\ \bibinfo {author} {\bibfnamefont {L.}~\bibnamefont {Sheng}},\
  }\href {\doibase 10.1103/PhysRevB.103.245409} {\bibfield  {journal} {\bibinfo
   {journal} {Phys. Rev. B}\ }\textbf {\bibinfo {volume} {103}},\ \bibinfo
  {pages} {245409} (\bibinfo {year} {2021})}\BibitemShut {NoStop}%
\bibitem [{\citenamefont {Lu}(2018)}]{lu20193d}%
  \BibitemOpen
  \bibfield  {author} {\bibinfo {author} {\bibfnamefont {H.-Z.}\ \bibnamefont
  {Lu}},\ }\href {\doibase 10.1093/nsr/nwy082} {\bibfield  {journal} {\bibinfo
  {journal} {National Science Review}\ }\textbf {\bibinfo {volume} {6}},\
  \bibinfo {pages} {208} (\bibinfo {year} {2018})}\BibitemShut {NoStop}%
\bibitem [{\citenamefont {Li}\ \emph {et~al.}(2021)\citenamefont {Li},
  \citenamefont {Wang}, \citenamefont {Du}, \citenamefont {Qin}, \citenamefont
  {Lu},\ and\ \citenamefont {Xie}}]{li20213d}%
  \BibitemOpen
  \bibfield  {author} {\bibinfo {author} {\bibfnamefont {S.}~\bibnamefont
  {Li}}, \bibinfo {author} {\bibfnamefont {C.}~\bibnamefont {Wang}}, \bibinfo
  {author} {\bibfnamefont {Z.}~\bibnamefont {Du}}, \bibinfo {author}
  {\bibfnamefont {F.}~\bibnamefont {Qin}}, \bibinfo {author} {\bibfnamefont
  {H.-Z.}\ \bibnamefont {Lu}}, \ and\ \bibinfo {author} {\bibfnamefont
  {X.}~\bibnamefont {Xie}},\ }\href {\doibase 10.1038/s41535-021-00399-2}
  {\bibfield  {journal} {\bibinfo  {journal} {npj Quantum Materials}\ }\textbf
  {\bibinfo {volume} {6}},\ \bibinfo {pages} {96} (\bibinfo {year}
  {2021})}\BibitemShut {NoStop}%
\bibitem [{\citenamefont {Qin}\ \emph {et~al.}(2020)\citenamefont {Qin},
  \citenamefont {Li}, \citenamefont {Du}, \citenamefont {Wang}, \citenamefont
  {Zhang}, \citenamefont {Yu}, \citenamefont {Lu},\ and\ \citenamefont
  {Xie}}]{PhysRevLett.125.206601}%
  \BibitemOpen
  \bibfield  {author} {\bibinfo {author} {\bibfnamefont {F.}~\bibnamefont
  {Qin}}, \bibinfo {author} {\bibfnamefont {S.}~\bibnamefont {Li}}, \bibinfo
  {author} {\bibfnamefont {Z.~Z.}\ \bibnamefont {Du}}, \bibinfo {author}
  {\bibfnamefont {C.~M.}\ \bibnamefont {Wang}}, \bibinfo {author}
  {\bibfnamefont {W.}~\bibnamefont {Zhang}}, \bibinfo {author} {\bibfnamefont
  {D.}~\bibnamefont {Yu}}, \bibinfo {author} {\bibfnamefont {H.-Z.}\
  \bibnamefont {Lu}}, \ and\ \bibinfo {author} {\bibfnamefont {X.~C.}\
  \bibnamefont {Xie}},\ }\href {\doibase 10.1103/PhysRevLett.125.206601}
  {\bibfield  {journal} {\bibinfo  {journal} {Phys. Rev. Lett.}\ }\textbf
  {\bibinfo {volume} {125}},\ \bibinfo {pages} {206601} (\bibinfo {year}
  {2020})}\BibitemShut {NoStop}%
\bibitem [{\citenamefont {Gooth}\ \emph {et~al.}(2023)\citenamefont {Gooth},
  \citenamefont {Galeski},\ and\ \citenamefont {Meng}}]{Gooth_2023}%
  \BibitemOpen
  \bibfield  {author} {\bibinfo {author} {\bibfnamefont {J.}~\bibnamefont
  {Gooth}}, \bibinfo {author} {\bibfnamefont {S.}~\bibnamefont {Galeski}}, \
  and\ \bibinfo {author} {\bibfnamefont {T.}~\bibnamefont {Meng}},\ }\href
  {\doibase 10.1088/1361-6633/acb8c9} {\bibfield  {journal} {\bibinfo
  {journal} {Reports on Progress in Physics}\ }\textbf {\bibinfo {volume}
  {86}},\ \bibinfo {pages} {044501} (\bibinfo {year} {2023})}\BibitemShut
  {NoStop}%
\bibitem [{\citenamefont {Wang}\ and\ \citenamefont
  {Cai}(2023)}]{PhysRevB.107.125203}%
  \BibitemOpen
  \bibfield  {author} {\bibinfo {author} {\bibfnamefont {Y.-X.}\ \bibnamefont
  {Wang}}\ and\ \bibinfo {author} {\bibfnamefont {Z.}~\bibnamefont {Cai}},\
  }\href {\doibase 10.1103/PhysRevB.107.125203} {\bibfield  {journal} {\bibinfo
   {journal} {Phys. Rev. B}\ }\textbf {\bibinfo {volume} {107}},\ \bibinfo
  {pages} {125203} (\bibinfo {year} {2023})}\BibitemShut {NoStop}%
\bibitem [{\citenamefont {Du}\ \emph {et~al.}(2021)\citenamefont {Du},
  \citenamefont {Wang}, \citenamefont {Sun}, \citenamefont {Lu},\ and\
  \citenamefont {Xie}}]{du2021quantum}%
  \BibitemOpen
  \bibfield  {author} {\bibinfo {author} {\bibfnamefont {Z.}~\bibnamefont
  {Du}}, \bibinfo {author} {\bibfnamefont {C.}~\bibnamefont {Wang}}, \bibinfo
  {author} {\bibfnamefont {H.-P.}\ \bibnamefont {Sun}}, \bibinfo {author}
  {\bibfnamefont {H.-Z.}\ \bibnamefont {Lu}}, \ and\ \bibinfo {author}
  {\bibfnamefont {X.}~\bibnamefont {Xie}},\ }\href {\doibase
  10.1038/s41467-021-25273-4} {\bibfield  {journal} {\bibinfo  {journal}
  {Nature communications}\ }\textbf {\bibinfo {volume} {12}},\ \bibinfo {pages}
  {5038} (\bibinfo {year} {2021})}\BibitemShut {NoStop}%
\bibitem [{\citenamefont {Uchida}\ \emph {et~al.}(2017)\citenamefont {Uchida},
  \citenamefont {Nakazawa}, \citenamefont {Nishihaya}, \citenamefont {Akiba},
  \citenamefont {Kriener}, \citenamefont {Kozuka}, \citenamefont {Miyake},
  \citenamefont {Taguchi}, \citenamefont {Tokunaga}, \citenamefont {Nagaosa}
  \emph {et~al.}}]{uchida2017quantum}%
  \BibitemOpen
  \bibfield  {author} {\bibinfo {author} {\bibfnamefont {M.}~\bibnamefont
  {Uchida}}, \bibinfo {author} {\bibfnamefont {Y.}~\bibnamefont {Nakazawa}},
  \bibinfo {author} {\bibfnamefont {S.}~\bibnamefont {Nishihaya}}, \bibinfo
  {author} {\bibfnamefont {K.}~\bibnamefont {Akiba}}, \bibinfo {author}
  {\bibfnamefont {M.}~\bibnamefont {Kriener}}, \bibinfo {author} {\bibfnamefont
  {Y.}~\bibnamefont {Kozuka}}, \bibinfo {author} {\bibfnamefont
  {A.}~\bibnamefont {Miyake}}, \bibinfo {author} {\bibfnamefont
  {Y.}~\bibnamefont {Taguchi}}, \bibinfo {author} {\bibfnamefont
  {M.}~\bibnamefont {Tokunaga}}, \bibinfo {author} {\bibfnamefont
  {N.}~\bibnamefont {Nagaosa}},  \emph {et~al.},\ }\href {\doibase
  10.1038/s41467-017-02423-1} {\bibfield  {journal} {\bibinfo  {journal}
  {Nature communications}\ }\textbf {\bibinfo {volume} {8}},\ \bibinfo {pages}
  {2274} (\bibinfo {year} {2017})}\BibitemShut {NoStop}%
\bibitem [{\citenamefont {Schumann}\ \emph {et~al.}(2018)\citenamefont
  {Schumann}, \citenamefont {Galletti}, \citenamefont {Kealhofer},
  \citenamefont {Kim}, \citenamefont {Goyal},\ and\ \citenamefont
  {Stemmer}}]{PhysRevLett.120.016801}%
  \BibitemOpen
  \bibfield  {author} {\bibinfo {author} {\bibfnamefont {T.}~\bibnamefont
  {Schumann}}, \bibinfo {author} {\bibfnamefont {L.}~\bibnamefont {Galletti}},
  \bibinfo {author} {\bibfnamefont {D.~A.}\ \bibnamefont {Kealhofer}}, \bibinfo
  {author} {\bibfnamefont {H.}~\bibnamefont {Kim}}, \bibinfo {author}
  {\bibfnamefont {M.}~\bibnamefont {Goyal}}, \ and\ \bibinfo {author}
  {\bibfnamefont {S.}~\bibnamefont {Stemmer}},\ }\href {\doibase
  10.1103/PhysRevLett.120.016801} {\bibfield  {journal} {\bibinfo  {journal}
  {Phys. Rev. Lett.}\ }\textbf {\bibinfo {volume} {120}},\ \bibinfo {pages}
  {016801} (\bibinfo {year} {2018})}\BibitemShut {NoStop}%
\bibitem [{\citenamefont {Zhang}\ \emph {et~al.}(2019)\citenamefont {Zhang},
  \citenamefont {Zhang}, \citenamefont {Yuan}, \citenamefont {Lu},
  \citenamefont {Zhang}, \citenamefont {Narayan}, \citenamefont {Liu},
  \citenamefont {Zhang}, \citenamefont {Ni}, \citenamefont {Liu} \emph
  {et~al.}}]{zhang2019quantum}%
  \BibitemOpen
  \bibfield  {author} {\bibinfo {author} {\bibfnamefont {C.}~\bibnamefont
  {Zhang}}, \bibinfo {author} {\bibfnamefont {Y.}~\bibnamefont {Zhang}},
  \bibinfo {author} {\bibfnamefont {X.}~\bibnamefont {Yuan}}, \bibinfo {author}
  {\bibfnamefont {S.}~\bibnamefont {Lu}}, \bibinfo {author} {\bibfnamefont
  {J.}~\bibnamefont {Zhang}}, \bibinfo {author} {\bibfnamefont
  {A.}~\bibnamefont {Narayan}}, \bibinfo {author} {\bibfnamefont
  {Y.}~\bibnamefont {Liu}}, \bibinfo {author} {\bibfnamefont {H.}~\bibnamefont
  {Zhang}}, \bibinfo {author} {\bibfnamefont {Z.}~\bibnamefont {Ni}}, \bibinfo
  {author} {\bibfnamefont {R.}~\bibnamefont {Liu}},  \emph {et~al.},\ }\href
  {\doibase 10.1038/s41586-018-0798-3} {\bibfield  {journal} {\bibinfo
  {journal} {Nature}\ }\textbf {\bibinfo {volume} {565}},\ \bibinfo {pages}
  {331} (\bibinfo {year} {2019})}\BibitemShut {NoStop}%
\bibitem [{\citenamefont {Tang}\ \emph {et~al.}(2019)\citenamefont {Tang},
  \citenamefont {Ren}, \citenamefont {Wang}, \citenamefont {Zhong},
  \citenamefont {Schneeloch}, \citenamefont {Yang}, \citenamefont {Yang},
  \citenamefont {Lee}, \citenamefont {Gu}, \citenamefont {Qiao} \emph
  {et~al.}}]{tang2019three}%
  \BibitemOpen
  \bibfield  {author} {\bibinfo {author} {\bibfnamefont {F.}~\bibnamefont
  {Tang}}, \bibinfo {author} {\bibfnamefont {Y.}~\bibnamefont {Ren}}, \bibinfo
  {author} {\bibfnamefont {P.}~\bibnamefont {Wang}}, \bibinfo {author}
  {\bibfnamefont {R.}~\bibnamefont {Zhong}}, \bibinfo {author} {\bibfnamefont
  {J.}~\bibnamefont {Schneeloch}}, \bibinfo {author} {\bibfnamefont {S.~A.}\
  \bibnamefont {Yang}}, \bibinfo {author} {\bibfnamefont {K.}~\bibnamefont
  {Yang}}, \bibinfo {author} {\bibfnamefont {P.~A.}\ \bibnamefont {Lee}},
  \bibinfo {author} {\bibfnamefont {G.}~\bibnamefont {Gu}}, \bibinfo {author}
  {\bibfnamefont {Z.}~\bibnamefont {Qiao}},  \emph {et~al.},\ }\href {\doibase
  10.1038/s41586-019-1180-9} {\bibfield  {journal} {\bibinfo  {journal}
  {Nature}\ }\textbf {\bibinfo {volume} {569}},\ \bibinfo {pages} {537}
  (\bibinfo {year} {2019})}\BibitemShut {NoStop}%
\bibitem [{\citenamefont {Lin}\ \emph {et~al.}(2019)\citenamefont {Lin},
  \citenamefont {Wang}, \citenamefont {Wiedmann}, \citenamefont {Lu},
  \citenamefont {Zheng}, \citenamefont {Yu},\ and\ \citenamefont
  {Liao}}]{PhysRevLett.122.036602}%
  \BibitemOpen
  \bibfield  {author} {\bibinfo {author} {\bibfnamefont {B.-C.}\ \bibnamefont
  {Lin}}, \bibinfo {author} {\bibfnamefont {S.}~\bibnamefont {Wang}}, \bibinfo
  {author} {\bibfnamefont {S.}~\bibnamefont {Wiedmann}}, \bibinfo {author}
  {\bibfnamefont {J.-M.}\ \bibnamefont {Lu}}, \bibinfo {author} {\bibfnamefont
  {W.-Z.}\ \bibnamefont {Zheng}}, \bibinfo {author} {\bibfnamefont
  {D.}~\bibnamefont {Yu}}, \ and\ \bibinfo {author} {\bibfnamefont {Z.-M.}\
  \bibnamefont {Liao}},\ }\href {\doibase 10.1103/PhysRevLett.122.036602}
  {\bibfield  {journal} {\bibinfo  {journal} {Phys. Rev. Lett.}\ }\textbf
  {\bibinfo {volume} {122}},\ \bibinfo {pages} {036602} (\bibinfo {year}
  {2019})}\BibitemShut {NoStop}%
\bibitem [{\citenamefont {Lifshitz}(1960)}]{lifshitz1960anomalies}%
  \BibitemOpen
  \bibfield  {author} {\bibinfo {author} {\bibfnamefont {I.}~\bibnamefont
  {Lifshitz}},\ }\href@noop {} {\bibfield  {journal} {\bibinfo  {journal} {Sov.
  Phys. JETP}\ }\textbf {\bibinfo {volume} {11}},\ \bibinfo {pages} {1130}
  (\bibinfo {year} {1960})}\BibitemShut {NoStop}%
\bibitem [{\citenamefont {Okada}\ \emph {et~al.}(2013)\citenamefont {Okada},
  \citenamefont {Serbyn}, \citenamefont {Lin}, \citenamefont {Walkup},
  \citenamefont {Zhou}, \citenamefont {Dhital}, \citenamefont {Neupane},
  \citenamefont {Xu}, \citenamefont {Wang}, \citenamefont {Sankar},
  \citenamefont {Chou}, \citenamefont {Bansil}, \citenamefont {Hasan},
  \citenamefont {Wilson}, \citenamefont {Fu},\ and\ \citenamefont
  {Madhavan}}]{science.1239451}%
  \BibitemOpen
  \bibfield  {author} {\bibinfo {author} {\bibfnamefont {Y.}~\bibnamefont
  {Okada}}, \bibinfo {author} {\bibfnamefont {M.}~\bibnamefont {Serbyn}},
  \bibinfo {author} {\bibfnamefont {H.}~\bibnamefont {Lin}}, \bibinfo {author}
  {\bibfnamefont {D.}~\bibnamefont {Walkup}}, \bibinfo {author} {\bibfnamefont
  {W.}~\bibnamefont {Zhou}}, \bibinfo {author} {\bibfnamefont {C.}~\bibnamefont
  {Dhital}}, \bibinfo {author} {\bibfnamefont {M.}~\bibnamefont {Neupane}},
  \bibinfo {author} {\bibfnamefont {S.}~\bibnamefont {Xu}}, \bibinfo {author}
  {\bibfnamefont {Y.~J.}\ \bibnamefont {Wang}}, \bibinfo {author}
  {\bibfnamefont {R.}~\bibnamefont {Sankar}}, \bibinfo {author} {\bibfnamefont
  {F.}~\bibnamefont {Chou}}, \bibinfo {author} {\bibfnamefont {A.}~\bibnamefont
  {Bansil}}, \bibinfo {author} {\bibfnamefont {M.~Z.}\ \bibnamefont {Hasan}},
  \bibinfo {author} {\bibfnamefont {S.~D.}\ \bibnamefont {Wilson}}, \bibinfo
  {author} {\bibfnamefont {L.}~\bibnamefont {Fu}}, \ and\ \bibinfo {author}
  {\bibfnamefont {V.}~\bibnamefont {Madhavan}},\ }\href {\doibase
  10.1126/science.1239451} {\bibfield  {journal} {\bibinfo  {journal}
  {Science}\ }\textbf {\bibinfo {volume} {341}},\ \bibinfo {pages} {1496}
  (\bibinfo {year} {2013})}\BibitemShut {NoStop}%
\bibitem [{\citenamefont {Zeljkovic}\ \emph {et~al.}(2014)\citenamefont
  {Zeljkovic}, \citenamefont {Okada}, \citenamefont {Huang}, \citenamefont
  {Sankar}, \citenamefont {Walkup}, \citenamefont {Zhou}, \citenamefont
  {Serbyn}, \citenamefont {Chou}, \citenamefont {Tsai}, \citenamefont {Lin},
  \citenamefont {Bansil}, \citenamefont {Fu}, \citenamefont {Hasan},\ and\
  \citenamefont {Madhavan}}]{Zeljkovic2014}%
  \BibitemOpen
  \bibfield  {author} {\bibinfo {author} {\bibfnamefont {I.}~\bibnamefont
  {Zeljkovic}}, \bibinfo {author} {\bibfnamefont {Y.}~\bibnamefont {Okada}},
  \bibinfo {author} {\bibfnamefont {C.-Y.}\ \bibnamefont {Huang}}, \bibinfo
  {author} {\bibfnamefont {R.}~\bibnamefont {Sankar}}, \bibinfo {author}
  {\bibfnamefont {D.}~\bibnamefont {Walkup}}, \bibinfo {author} {\bibfnamefont
  {W.}~\bibnamefont {Zhou}}, \bibinfo {author} {\bibfnamefont {M.}~\bibnamefont
  {Serbyn}}, \bibinfo {author} {\bibfnamefont {F.}~\bibnamefont {Chou}},
  \bibinfo {author} {\bibfnamefont {W.-F.}\ \bibnamefont {Tsai}}, \bibinfo
  {author} {\bibfnamefont {H.}~\bibnamefont {Lin}}, \bibinfo {author}
  {\bibfnamefont {A.}~\bibnamefont {Bansil}}, \bibinfo {author} {\bibfnamefont
  {L.}~\bibnamefont {Fu}}, \bibinfo {author} {\bibfnamefont {M.~Z.}\
  \bibnamefont {Hasan}}, \ and\ \bibinfo {author} {\bibfnamefont
  {V.}~\bibnamefont {Madhavan}},\ }\href {\doibase 10.1038/nphys3012}
  {\bibfield  {journal} {\bibinfo  {journal} {Nat. Phys.}\ }\textbf {\bibinfo
  {volume} {10}},\ \bibinfo {pages} {572} (\bibinfo {year} {2014})}\BibitemShut
  {NoStop}%
\bibitem [{\citenamefont {Wu}\ \emph {et~al.}(2015)\citenamefont {Wu},
  \citenamefont {Jo}, \citenamefont {Ochi}, \citenamefont {Huang},
  \citenamefont {Mou}, \citenamefont {Bud'ko}, \citenamefont {Canfield},
  \citenamefont {Trivedi}, \citenamefont {Arita},\ and\ \citenamefont
  {Kaminski}}]{PhysRevLett.115.166602}%
  \BibitemOpen
  \bibfield  {author} {\bibinfo {author} {\bibfnamefont {Y.}~\bibnamefont
  {Wu}}, \bibinfo {author} {\bibfnamefont {N.~H.}\ \bibnamefont {Jo}}, \bibinfo
  {author} {\bibfnamefont {M.}~\bibnamefont {Ochi}}, \bibinfo {author}
  {\bibfnamefont {L.}~\bibnamefont {Huang}}, \bibinfo {author} {\bibfnamefont
  {D.}~\bibnamefont {Mou}}, \bibinfo {author} {\bibfnamefont {S.~L.}\
  \bibnamefont {Bud'ko}}, \bibinfo {author} {\bibfnamefont {P.~C.}\
  \bibnamefont {Canfield}}, \bibinfo {author} {\bibfnamefont {N.}~\bibnamefont
  {Trivedi}}, \bibinfo {author} {\bibfnamefont {R.}~\bibnamefont {Arita}}, \
  and\ \bibinfo {author} {\bibfnamefont {A.}~\bibnamefont {Kaminski}},\ }\href
  {\doibase 10.1103/PhysRevLett.115.166602} {\bibfield  {journal} {\bibinfo
  {journal} {Phys. Rev. Lett.}\ }\textbf {\bibinfo {volume} {115}},\ \bibinfo
  {pages} {166602} (\bibinfo {year} {2015})}\BibitemShut {NoStop}%
\bibitem [{\citenamefont {Ding}\ \emph {et~al.}(2021)\citenamefont {Ding},
  \citenamefont {Zhu}, \citenamefont {Hu},\ and\ \citenamefont
  {Su}}]{PhysRevB.104.155135}%
  \BibitemOpen
  \bibfield  {author} {\bibinfo {author} {\bibfnamefont {K.-H.}\ \bibnamefont
  {Ding}}, \bibinfo {author} {\bibfnamefont {Z.-G.}\ \bibnamefont {Zhu}},
  \bibinfo {author} {\bibfnamefont {Y.-L.}\ \bibnamefont {Hu}}, \ and\ \bibinfo
  {author} {\bibfnamefont {G.}~\bibnamefont {Su}},\ }\href {\doibase
  10.1103/PhysRevB.104.155135} {\bibfield  {journal} {\bibinfo  {journal}
  {Phys. Rev. B}\ }\textbf {\bibinfo {volume} {104}},\ \bibinfo {pages}
  {155135} (\bibinfo {year} {2021})}\BibitemShut {NoStop}%
\end{thebibliography}%

\end{document}